\documentclass[11pt,draftcls,onecolumn]{IEEEtran}

\usepackage{graphicx}
\usepackage{amssymb,amsmath,dsfont,amsfonts}
\usepackage{epstopdf}
\usepackage{cite}
\usepackage{color}
\usepackage{boxedminipage}
\usepackage{listings}
\usepackage{subfigure}
\usepackage{caption2}
\usepackage{float}

\newtheorem{theorem}{Theorem}
\newtheorem{lemma}{Lemma}

\newtheorem{proposition}{Proposition}
\newtheorem{definition}{Definition}
\newtheorem{example}{Example}

\newtheorem{remark}{Remark}
\newtheorem{conjecture}{Conjecture}

\newcommand{\prf}[1]{{\bf Proof} \, #1 \hfill $\blacksquare$}

\newcommand{\cut}[1]{}

\newcommand{\vc}[1]{\boldsymbol{#1}}

\newcommand{\Rcal}{{\mathcal R}}

\def\real    { \mathbb{R} }

\newcommand{\Tr}[1]{{\rm Tr}\left( #1 \right)}
\newcommand{\norm}[1]{\| #1 \|}

\newcommand{\Prob}[1]{{\bf P}\left\{#1\right\}}
\newcommand{\E}[1]{{\bf E}\left[#1\right]}

\newcounter{l1}
\newcommand{\barablist}{\begin{list}{\arabic{l1}}{\usecounter{l1}}}

\usepackage{acronym}
\acrodef{i.i.d.}{independent and identically distributed}
\acrodef{LTI}{Linear Time-Invariant}
\acrodef{RIP}{Restricted Isometry Property}
\acrodef{SVD}{Singular Value Decomposition}
\acrodef{CS}{Compressive Sensing}
\acrodef{DSP}{Digital Signal Processing}

\acrodef{FIR}{Finite Impulse Response}
\acrodef{LTI}{linear time-invariant}
\acrodef{DFT}{Discrete Fourier Transform}
\acrodef{JL}{Johnson-Lindenstrauss}
\acrodef{ROC}{Receiver Operating Curve}
\acrodef{NP}{Neyman-Pearson}
\acrodef{CoM}{Concentration of Measure}
\acrodef{CSI}{Compressive System Identification}
\acrodef{CBD}{Compressive Binary Detection}

\begin{document}

\title{Concentration of Measure Inequalities for Toeplitz Matrices with Applications
}

\author{Borhan M. Sanandaji, Tyrone L. Vincent, and Michael B. Wakin
\thanks{All authors are with the Department of Electrical Engineering and Computer Science, Colorado School of Mines, Golden, CO 80401, USA. Email: \{bmolazem, tvincent, mwakin\}@mines.edu. This work was partially supported by AFOSR Grant FA9550-09-1-0465, NSF Grant CCF-0830320, DARPA Grant HR0011-08-1-0078, and NSF Grant CNS-0931748.}
}

\maketitle

\vspace{-0.5in}

\begin{abstract}
We derive \ac{CoM} inequalities for randomized Toeplitz matrices.
These inequalities show that the norm of a high-dimensional signal mapped by a Toeplitz matrix to a low-dimensional space concentrates around its mean with a tail probability bound that decays exponentially in the dimension of the range space divided by a quantity which is a function of the signal.
For the class of {\em sparse} signals, the introduced quantity is bounded by the sparsity level of the signal. However, we observe that this bound is highly pessimistic for most sparse signals and we show that if a random distribution is imposed on the non-zero entries of the signal, the typical value of the quantity is bounded by a term that scales logarithmically in the ambient dimension.
As an application of the \ac{CoM} inequalities, we consider \ac{CBD}.
\end{abstract}


\acresetall
\section{Introduction}
\label{sec:intro}

\subsection{Overview}
\label{sec:overview}

\IEEEPARstart{M}{otivated} to reduce the burdens of acquiring, transmitting, storing, and analyzing vast quantities of data, signal processing researchers have over the last few decades developed a variety of techniques for data compression and dimensionality reduction.
Unfortunately, many of these techniques require a raw, high-dimensional data set to be acquired before its essential low-dimensional structure can be identified, extracted, and exploited.
In contrast, what would be truly desirable are sensors/operators that require fewer raw measurements yet still capture the essential information in a data set. These operators can be called {\em compressive} in the sense that they act as mappings from a high-dimensional to a low-dimensional space, e.g., $X: \real^N \to \real^M$ where $M < N$. Linear compressive operators correspond to matrices having fewer rows than columns. Although such matrices can have arbitrary/deterministic entries, {\em randomized} matrices (those with entries drawn from a random distribution) have attracted the attention of researchers due to their universality and ease of analysis. Utilizing such compressive operators to achieve {\em information-preserving embeddings} of high-dimensional (but compressible) data sets into low-dimensional spaces can drastically simplify the acquisition process, reduce the needed amount of storage space, and decrease the computational demands of data processing.

\ac{CoM} inequalities are one of the leading techniques used in the theoretical analysis of randomized compressive linear operators~\cite{ledoux2001concentration,dasgupta2003elementary,achlioptas2003database}. These inequalities quantify how well a random matrix will preserve the norm of a high-dimensional signal when mapping it to a low-dimensional space.
 A typical \ac{CoM} inequality takes the following form. For any fixed signal $\vc{a} \in \real^{N}$, and a suitable random $M \times N$ matrix $X$,
the random variable $\|X\vc{a}\|_2^2$ will be highly concentrated around its expected value, $\E{\|X\vc{a}\|_2^2}$, with high probability. Formally, there exist constants $c_1$ and $c_2$ such that for any fixed $\vc{a} \in \real^{N}$,
\begin{equation}
\Prob{\left|\|X\vc{a}\|_2^2-\E{\|X\vc{a}\|_2^2}\right|  \geq \epsilon \E{\|X\vc{a}\|_2^2}} \leq c_1 e^{-c_2Mc_0\left(\epsilon\right)},
\label{eq:com_uniform1}
\end{equation}
where $c_0\left(\epsilon\right)$ is a positive constant that depends on $\epsilon \in (0,1)$.

\ac{CoM} inequalities for random operators have been shown to have important implications in signal processing and machine learning. One of the most prominent results in this area is the \ac{JL} lemma, which concerns embedding a finite set of points in a lower dimensional space using a distance preserving mapping~\cite{johnson1984extensions}.
Dasgupta et al.~\cite{dasgupta2003elementary} and Achlioptas~\cite{achlioptas2003database} showed how a \ac{CoM} inequality of the form~(\ref{eq:com_uniform1}) could establish that with high probability, an \ac{i.i.d.} random compressive operator $X \in \real^{M \times N}$ ($M < N$) provides a \ac{JL}-embedding.
Specifically, for a given $\epsilon \in (0,1)$, for any fixed point set $Q \subseteq \real^N$,
\begin{equation}
\left(1-\epsilon\right)\|\vc{a}-\vc{b}\|_2^2 \leq \|X\vc{a}-X\vc{b}\|_2^2 \leq \left(1+\epsilon\right)\|\vc{a}-\vc{b}\|_2^2
\label{jllemma}
\end{equation}
holds with high probability for all $\vc{a}$, $\vc{b} \in Q$ if $M = \mathcal{O}\left(\epsilon^{-2}\text{log}\left(|Q|\right)\right)$.
One of the other significant consequences of \ac{CoM} inequalities is in the context of \ac{CS}~\cite{candes2008people} and the \ac{RIP}~\cite{candes2005decoding}. If a matrix $X$ satisfies (\ref{jllemma}) for all pairs $\vc{a}$, $\vc{b}$ of $K$-sparse signals in $\real^{N}$, then $X$ is said to satisfy the \ac{RIP} of order $2K$ with isometry constant $\epsilon$. 
Establishing the \ac{RIP} of order $2K$ for a given compressive matrix $X$ 
leads to understanding the number of measurements required to have exact recovery for any $K$-sparse signal
$\vc{a} \in \real^N$. Baraniuk et al.~\cite{baraniuk2008simple} and Mendelson et al.~\cite{mendelson2008uniform} showed that \ac{CoM} inequalities can be used to prove the \ac{RIP} for random compressive matrices.

\ac{CoM} inequalities have been well-studied and derived for {\em unstructured} random compressive matrices, populated with \ac{i.i.d.} random entries \cite{achlioptas2003database,dasgupta2003elementary}. However, in many practical applications, measurement matrices possess a certain structure. In particular, when linear dynamical systems are involved, Toeplitz and circulant matrices appear due to the convolution process~\cite{rauhut9circulant,haupt-toeplitz,romberg2008compressive,sanandaji2010toeplitz}. Specifically, consider the \ac{LTI} dynamical system with system finite impulse response $\vc{a} = \{a_{k}\}_{k=1}^{N}$. Let $\vc{x} = \{x_{k}\}_{k=1}^{N+M-1}$ be the applied input sequence. Then the corresponding output is calculated from the time-domain convolution of $\vc{a}$ and $\vc{x}$. 
Supposing the $x_k$ and $a_k$ sequences are zero-padded from both sides, each output sample $y_k$ can be written as
\begin{equation}
y_k = \sum_{j=1}^{N}a_jx_{k-j}.
\label{firfilter}
\end{equation}
If we keep only $M$ consecutive observations of the system, $\vc{y} = \{y_{k}\}_{k=N+1}^{N+M}$, then~(\ref{firfilter}) can be written in matrix-vector multiplication format as
\begin{equation}
\vc{y} = X\vc{a},
\label{matrixmulti}
\end{equation}
where
\begin{equation}
X = \left[ \begin{array}{cccc} x_N & x_{N-1} & \cdots & x_1 \\
x_{N+1} & x_N & \cdots & x_2 \\ \vdots & \vdots & \ddots & \vdots \\
x_{N+M-1} & x_{N+M-2} & \cdots & x_M \end{array} \right]
\label{toeplitzmatrix}
\end{equation}
is an $M \times N$ Toeplitz matrix. If the entries of $X$ are generated randomly, we say $X$ is a randomized Toeplitz matrix.
Other types of {\em structured} random matrices also arise when dynamical systems are involved. For example block-diagonal matrices appear in applications such as distributed sensing systems~\cite{park2011block} and initial state estimation (observability) of linear systems~\cite{wakin2010observability}. 

In this paper, we consider compressive randomized Toeplitz matrices, derive \ac{CoM} inequalities, and discuss their implications in applications such as sparse impulse response recovery~\cite{bajwa8compressed,sanandaji2011csi,haupt-toeplitz}. We also consider the problem of detecting a deviation in a system's behavior. We show that by characterizing the deviation using a particular measure that appears in our \ac{CoM} inequality, the detector performance can be correctly predicted.


%
%
%
%

\subsection{Related Work}
Compressive Toeplitz (and circulant) matrices have been previously studied in the context of \ac{CS}~\cite{tropp2006random,bajwa2007toeplitz,romberg2008compressive,bajwa8compressed,haupt-toeplitz,rauhut9circulant}, with applications involving 
channel estimation, synthetic aperture radar, etc.
Tropp et al.~\cite{tropp2006random} originally considered compressive Toeplitz matrices in an early \ac{CS} paper that proposed an efficient measurement mechanism involving a \ac{FIR} filter with random taps. 
Motivated by applications related to sparse channel estimation, Bajwa et al.~\cite{bajwa2007toeplitz} studied such matrices more formally in the case where the matrix entries are drawn from a symmetric Bernoulli distribution. Later they extended this study to random matrices whose entries are bounded or Gaussian-distributed and showed that with high probability, $M \geq \mathcal{O}\left(K^2\log\left(\frac{N}{K}\right)\right)$ measurements are sufficient to establish the \ac{RIP} of order $2K$ for vectors sparse in the time domain~\cite{bajwa8compressed,haupt-toeplitz}.
(It should be noted that the quadratic \ac{RIP} result can also be achieved using other methods such as a coherence argument~\cite{rauhut9circulant,rauhut2010compressive}.)
Recently, using more complicated mathematical tools such as Dudley's inequality for chaos and generic chaining, Rauhut et al.~\cite{rauhut2010partial} showed that with
$M \geq \mathcal{O}\left(K^{1.5}\log\left(N\right)^{1.5}\right)$ measurements the \ac{RIP} of order $2K$ will hold.\footnote{After the initial submission of our manuscript, in their very recent work, Krahmer et al.~\cite{rauhut2012suprema} showed that the minimal required number of measurements scales linearly with $K$, or formally $M \geq \mathcal{O}\left(K\log\left(K\right)^{2}\log\left(N\right)^{2}\right)$ measurements are sufficient to establish the \ac{RIP} of order $2K$. The recent linear \ac{RIP} result confirms what is suggested by simulations.} 
Note that these bounds compare to $M \geq \mathcal{O}\left(K\log\left(\frac{N}{K}\right)\right)$ measurements which are known to suffice when $X$ is unstructured~\cite{baraniuk2008simple}. 

In this paper, we derive \ac{CoM} inequalities for Toeplitz matrices and show how these inequalities reveal non-uniformity and signal-dependency of the mappings.
As one consequence of these \ac{CoM} inequalities, one could use them (along with standard covering number estimates) to prove the \ac{RIP} for compressive Toeplitz matrices. Although the estimate of the required number of measurements would be quadratic in terms of sparsity (i.e., $M \sim K^2$) 
and fall short of the best known estimates described above, studying concentration inequalities for Toeplitz matrices is of its own interest and gives insight to other applications such as the binary detection problem. 

There also exist \ac{CoM} analyses for other types of structured matrices.
For example, Park et al.~\cite{park2011block} derived concentration bounds for two types of block diagonal compressive matrices, one in which the blocks along the diagonal are random and independent, and one in which the blocks are random but equal.\footnote{Shortly after our own development of \ac{CoM} inequalities for compressive Toeplitz matrices (a preliminary version of Theorem~\ref{maintheo1} appeared in~\cite{sanandaji2010toeplitz}), Yap and Rozell~\cite{yapTechRepToep} showed that similar inequalities can be derived by extending the \ac{CoM} results for block diagonal matrices. Our Theorem~\ref{maintheo2} and the associated discussion, however, is unique to this paper.}
We subsequently extended these \ac{CoM} results for block diagonal matrices to the observability matrices that arise in the analysis of linear dynamical systems~\cite{wakin2010observability}.

\subsection{Contributions}
In summary, we derive \ac{CoM} inequalities for randomized Toeplitz matrices.
The derived bounds in the inequalities are non-uniform and depend on a quantity which is a function of the signal.
For the class of {\em sparse} signals, the introduced quantity is bounded by the sparsity level of the signal while if a random distribution is imposed on the non-zero entries of the signal, the typical value of the quantity is bounded by a term that scales logarithmically in the ambient dimension. As an application of the \ac{CoM} inequalities, we consider \ac{CBD}.

\section{Main Results}
\label{sec:main_result}
In this paper, we derive \ac{CoM} bounds for compressive Toeplitz matrices as given in~(\ref{toeplitzmatrix}) with entries $\{x_{k}\}_{k=1}^{N+M-1}$ drawn from an \ac{i.i.d.} Gaussian random sequence. Our first main result, detailed in Theorem~\ref{maintheo1}, states that the upper and lower tail probability bounds depend on the number of measurements $M$ and on the eigenvalues of 
the covariance matrix of the vector $\vc{a}$ defined as
\begin{equation}
P\left(\vc{a}\right) = \begin{bmatrix} \Rcal_{\vc{a}}\left(0\right) & \Rcal_{\vc{a}}\left(1\right) & \cdots & \Rcal_{\vc{a}}\left(M-1\right) \\
\Rcal_{\vc{a}}\left(1\right) & \Rcal_{\vc{a}}\left(0\right) & \cdots & \Rcal_{\vc{a}}\left(M-2\right) \\
\vdots & \vdots & \ddots & \vdots \\
\Rcal_{\vc{a}}\left(M-1\right) & \Rcal_{\vc{a}}\left(M-2\right) & \cdots & \Rcal_{\vc{a}}\left(0\right)
\end{bmatrix},
\label{covmat1}
\end{equation}
where
$\mathcal{R}_{\vc{a}}\left(\tau\right) := \sum_{i=1}^{N-\tau} a_i a_{i+\tau}$
denotes the un-normalized sample autocorrelation function of $\vc{a} \in \real^{N}$.
\begin{theorem}
Let $\vc{a} \in \real^N$ be fixed. Define two quantities $\rho\left(\vc{a}\right)$ and $\mu\left(\vc{a}\right)$ associated with the eigenvalues of the covariance matrix $P\left(\vc{a}\right)$ as
$
\rho\left(\vc{a}\right) := \frac{\max_i \lambda_{i}}{\|\vc{a}\|_2^2}
$
and
$
\mu\left(\vc{a}\right) := \frac{\sum_{i=1}^{M}\lambda_i^2}{M \|\vc{a}\|_2^4},
$
where $\lambda_{i}$ is the $i$-th eigenvalue of $P\left(\vc{a}\right)$. Let $\vc{y} = X\vc{a}$, where $X$ is a random compressive Toeplitz matrix with \ac{i.i.d.} Gaussian entries having zero mean and unit variance. Noting that $\E{\|\vc{y}\|_2^2} = M\|\vc{a}\|_2^2$, then for any $\epsilon \in \left(0,1\right)$, the upper tail probability bound is
\begin{equation}
\Prob{\|\vc{y}\|_2^{2} - M\|\vc{a}\|_2^{2} \geq \epsilon M\|\vc{a}\|_2^{2}} \leq e^{-\frac{\epsilon^{2}M}{8\rho\left(\vc{a}\right)}}
\label{eq1}
\end{equation}
and the lower tail probability bound is
\begin{equation}
\Prob{\|\vc{y}\|_2^{2} - M\|\vc{a}\|_2^{2} \leq -\epsilon M\|\vc{a}\|_2^{2}} \leq e^{-\frac{\epsilon^{2}M}{8\mu\left(\vc{a}\right)}}.
\label{eq2}
\end{equation}
\label{maintheo1}
\end{theorem}
Theorem~\ref{maintheo1} provides \ac{CoM} inequalities for {\em any} (not necessarily sparse) signal $\vc{a} \in \real^N$. The significance of these 
results comes from the fact that the tail probability bounds are functions 
of the signal $\vc{a}$, where the dependency is captured in the 
quantities $\rho\left(\vc{a}\right)$ and $\mu\left(\vc{a}\right)$.
This is not the case when $X$ is unstructured. Indeed, allowing $X$ to have $M\times N$ i.i.d.\ Gaussian entries with zero mean and unit variance (and thus, no Toeplitz structure) would result in the concentration inequality (see, e.g., \cite{achlioptas2003database})
\begin{equation}
\Prob{\left|\|\vc{y}\|_2^{2} - M\|\vc{a}\|_2^{2}\right| \geq \epsilon M\|\vc{a}\|_2^{2}} \leq 2e^{-\frac{\epsilon^{2}M}{4}}.
\label{eq:conc_unstruct}
\end{equation}
Thus, comparing the bound in (\ref{eq:conc_unstruct}) with the ones in (\ref{eq1}) and (\ref{eq2}), one could conclude that achieving the same probability bound for Toeplitz matrices requires choosing $M$ larger by a factor of $2\rho\left(\vc{a}\right)$ or $2\mu\left(\vc{a}\right)$.
Typically, when using \ac{CoM} inequalities such as~\eqref{eq1} and~\eqref{eq2}, we must set $M$ large enough so that both bounds are sufficiently small over all signals $\vc{a}$ belonging to some class of interest. For example, we are often interested in signals that have a {\em sparse} representation. Because we generally wish to keep $M$ as small as possible, it is interesting to try to obtain an upper bound for the important quantities $\rho\left(\vc{a}\right)$ and $\mu\left(\vc{a}\right)$ over the class of signals of interest. It is easy to show that for all $\vc{a} \in \real^N$, $\mu\left(\vc{a}\right) \leq \rho\left(\vc{a}\right)$. Thus, we limit our analysis to finding the sharpest upper bound for $\rho\left(\vc{a}\right)$ when $\vc{a}$ is $K$-sparse. For the sake of generality, we allow the signal to be sparse in an arbitrary orthobasis.

\begin{definition}
A signal $\vc{a} \in \real^N$ is called {\em $K$-sparse in an orthobasis $G \in \real^{N \times N}$} if it can be represented as $\vc{a} = G\vc{q}$,
where $\vc{q} \in \real^N$ is $K$-sparse (a vector with $K < N$ non-zero entries).
\label{def:basis1}
\end{definition}

We also introduce the notion of {\em $K$-sparse Fourier coherence} of 
the orthobasis $G$. This measures how strongly the columns of $G$ are correlated with the length $L$ Fourier basis, $F_{L}\in \mathbb{C}^{L\times L}$, which has entries $F_{L}(\ell,m) =\frac{1}{\sqrt{L}}w^{(\ell-1)(m-1)}$, where $w=e^{-\frac{2\pi j}{L}}$.

\begin{definition} Given an orthobasis $G \in \real^{N \times N}$ and measurement length $M$, let $L=N+M-1$. The {\em $K$-sparse Fourier coherence of $G$}, denoted $\nu_K\left(G\right)$, is defined as 
\begin{equation}
\nu_K\left(G\right) := \max_{i, S} \|F^{i\to}_{1:N} G_S\|_2,
\label{eq:nu_def}
\end{equation}
where $S \subseteq \{1, 2, \dots, N\}$ is the {\em support set} and varies over all possible sets with cardinality $\left|{S}\right| = K$, $G_S \in \real^{N \times K}$ is a matrix containing the columns of $G \in \real^{N \times N}$ indexed by the support set $S$, and $F^{i\to}_{1:N}\in \mathbb{C}^{N}$ is a row 
vector containing the first $N$ entries of the $i$-th row of the Fourier orthobasis $F_L \in \mathbb{C}^{L \times L}$. Observe that for a given orthobasis $G$,
$\nu_K\left(G\right)$ depends on $K$.
\label{def:nu1}
\end{definition}

Using the notion of Fourier coherence, we show in 
Section~\ref{sec:theo2} that for all vectors $\vc{a} \in \real^N$ that are $K$-sparse in an orthobasis $G \in \real^{N \times N}$,
\begin{equation}
\rho \left(\vc{a}\right) \leq L\nu_K^2\left(G\right),
\label{eq:consbnd}
\end{equation}
where, as above, $L=N+M-1$. This bound, however, appears to be highly pessimistic for most $K$-sparse signals. As a step towards better understanding the behavior of $\rho \left(\vc{a}\right)$, we consider a random model for $\vc{a}$. In particular, we consider a fixed $K$-sparse support set, and on this set we suppose the $K$ non-zero entries of the coefficient vector $\vc{q}$ are drawn from a random distribution. Based on this model, we derive an upper bound for $\E{\rho \left(\vc{a}\right)}$.

\begin{theorem}{({\em Upper Bound on $\E{\rho\left(\vc{a}\right)}$})}
Let $\vc{q} \in \mathbb{R}^N$ be a random $K$-sparse vector whose $K$ non-zero entries (on an arbitrary support $S$) are i.i.d.\ random variables drawn from a Gaussian distribution with $\mathcal{N} (0, \frac{1}{K})$. Select 
the measurement length $M$, which corresponds to the dimension of $P(\vc{a})$, and set $L=N+M-1$. Let $\vc{a} = G\vc{q}$ where $G \in \real^{N \times N}$ is an orthobasis. Then
\begin{equation}
\E{\rho\left(\vc{a}\right)} \leq \frac{8L\nu_K^2\left(G\right)}{K}\left(\log{2L}+2\right).
\label{eq:expected_upper_bnd_1}
\end{equation}
\label{maintheo2}
\end{theorem}

%
The $K$-sparse Fourier coherence $\nu_K\left(G\right)$ and consequently the bounds (\ref{eq:consbnd}) and (\ref{eq:expected_upper_bnd_1}) can be explicitly evaluated for some specific orthobases $G$. For example, letting $G = I_N$ (the $N \times N$ identity matrix), we can consider signals that are sparse in the time domain. With this choice of $G$, one can show that $\nu_K\left(I_N\right) = \sqrt{\frac{K}{L}}$. 
As another example, we can consider signals that are sparse in the frequency domain. To do this, we set $G$ equal to a real-valued version of the Fourier orthobasis. Without loss of generality, suppose $N$ is even. The real Fourier orthobasis, denoted $R_N$, is constructed as follows. The first column of $R_N$ equals the first column of $F_N$. Then $R_{\{2, \dots, \frac{N}{2}\}} = \text{Real}\left(\sqrt{2}F_{\{2, \dots, \frac{N}{2}\}}\right)$ and $R_{\{\frac{N}{2} + 1, \dots, N-1\}} = \text{Imaginary}\left(\sqrt{2}F_{\{2, \dots, \frac{N}{2}\}}\right)$. The last column of $R_N$ is equal to the $(\frac{N}{2}+1)$-th column of $F_N$. Similar steps can be taken to construct a real Fourier orthobasis when $N$ is odd. With this choice of $G$, one can show that
$
\nu_K\left(R_N\right) \leq \sqrt{\frac{N}{L}}.
$
Using these upper bounds on the Fourier coherence, we have the 
following deterministic bounds on $\rho \left(\vc{a}\right)$ in the time and frequency domains:
\begin{align}
\rho \left(\vc{a}\right) &\leq K \ \ \left(\text{time domain sparsity}\right) \ \ \text{and} \label{eq:detTime}\\
\rho \left(\vc{a}\right) &\leq N \ \ \left(\text{frequency domain sparsity}\right). \label{eq:detFreq}
\end{align}
We also obtain the following bounds on the expected value of $\rho \left(\vc{a}\right)$ under the random signal model:
\begin{align}
\E{\rho\left(\vc{a}\right)} &\leq 8\left(\log{2L}+2\right) \ \ \left(\text{time domain sparsity}\right) \ \ \text{and}  \label{eq:probTime}\\
\E{\rho\left(\vc{a}\right)} &\leq \frac{8N}{K}\left(\log{2L}+2\right) \ \ \left(\text{frequency domain sparsity}\right). \label{eq:probFreq}
\end{align}

\begin{figure}[tb]
\centering
\subfigure[]{
   \includegraphics[width = .45\columnwidth]{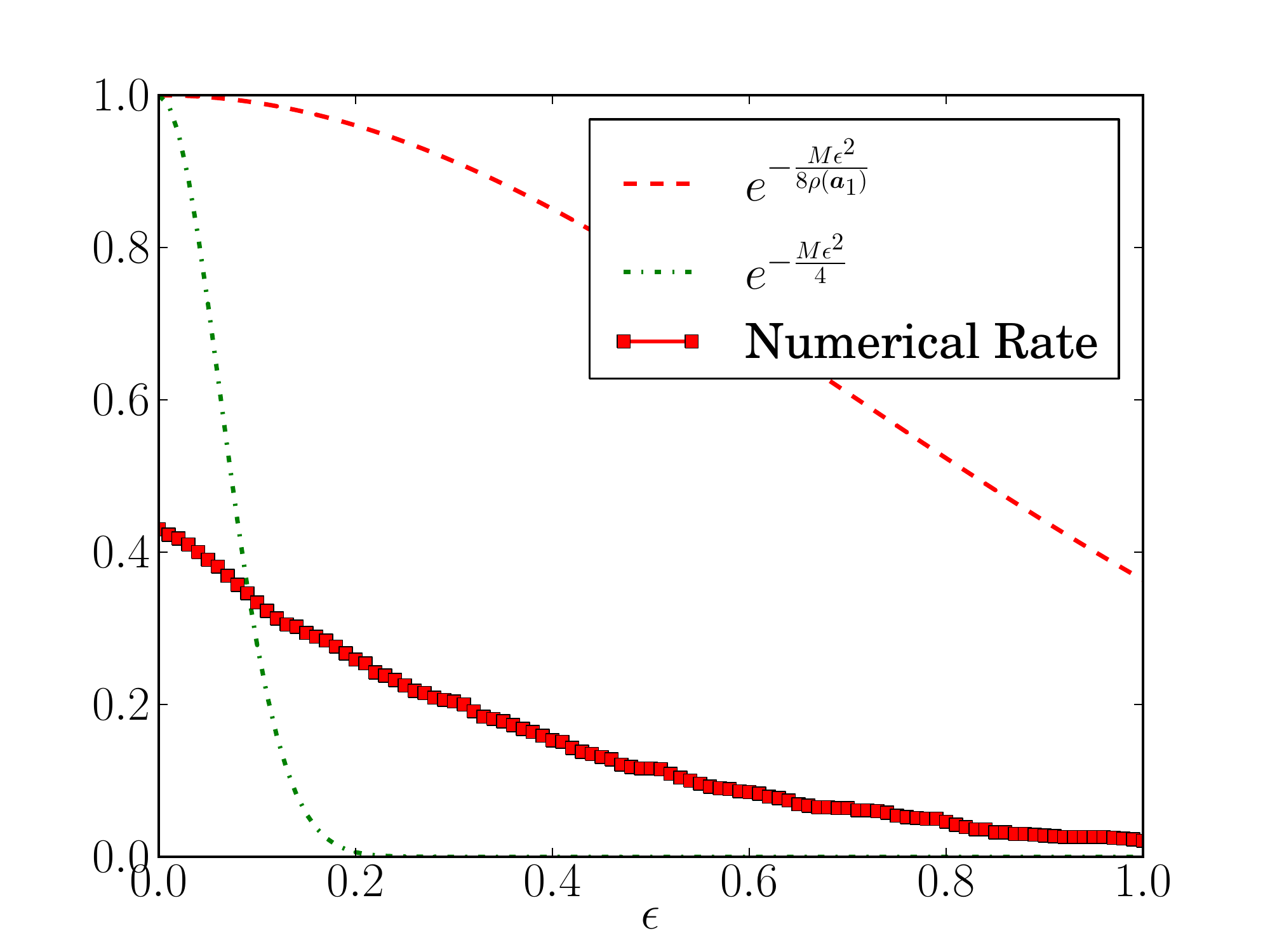}
   \label{fig:a1}
 }
\subfigure[]{
   \includegraphics[width = .45\columnwidth]{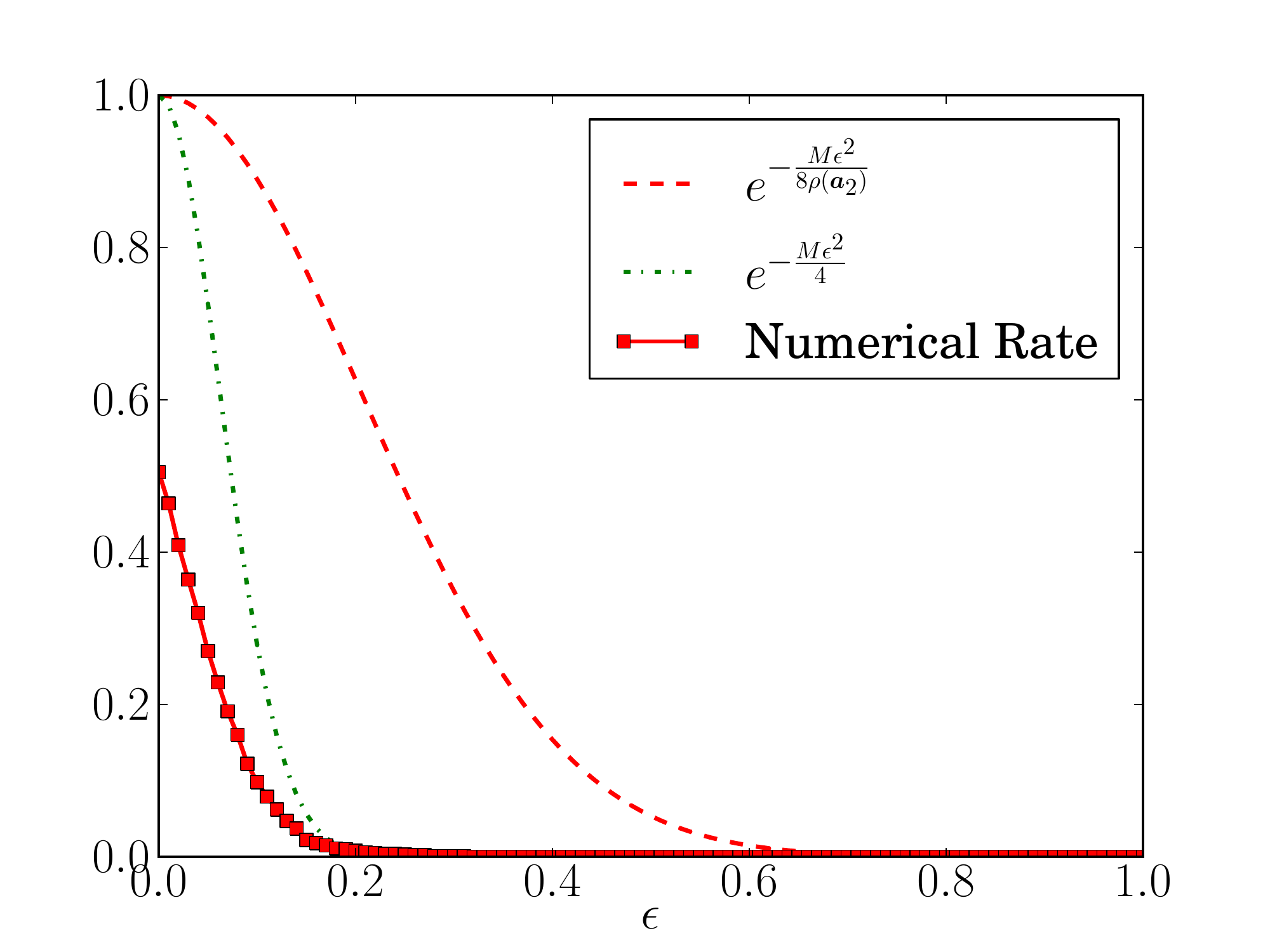}
   \label{fig:a2}
 }
\caption{\small \sl Illustrating the signal-dependency of the left-hand side of \ac{CoM} inequality~\eqref{eq1}. With fixed $N = 1024$, $K = 64$ and $M = 512$, we consider two particular $K$-sparse signals, $\vc{a_1}$, and $\vc{a_2}$. Both $\vc{a}_1$ and $\vc{a}_2$ are normalized. We measure each of these signals with $1000$ \ac{i.i.d.} Gaussian $M \times N$ Toeplitz matrices. (a) The $K$ non-zero entries of $\vc{a}_1$ have equal values and occur in the first $K$ entries of the vector. (b) The $K$ non-zero entries of $\vc{a}_2$ appear in randomly-selected locations with random signs and values. 
The two signals have different concentrations that can be upper bounded by the signal-dependent right-hand side of~\eqref{eq1}.
}
\label{fig:sig_dep_bnd}
\end{figure}

We offer a brief interpretation and analysis of these bounds in this paragraph and several examples that follow.
First, because $K \le N$, the deterministic and expectation bounds on $\rho \left(\vc{a}\right)$ are smaller for signals that are sparse in the time domain than for signals that are sparse in the frequency domain. 
The simulations described in Examples~\ref{exm:basis_256} and~\ref{exm:dft_1024} below confirm that, on average, $\rho \left(\vc{a}\right)$ does indeed tend to be smaller under the model of time domain sparsity.
Second, these bounds exhibit varying dependencies on the sparsity level $K$: \eqref{eq:detTime} increases with $K$ and \eqref{eq:probFreq} decreases with $K$, while \eqref{eq:detFreq} and \eqref{eq:probTime} are agnostic to $K$.
The simulation described in Example~\ref{exm:basis_256} below confirms that, on average, $\rho \left(\vc{a}\right)$ increases with $K$ for signals that are sparse in the time domain but decreases with $K$ for signals that are sparse in the frequency domain. This actually reveals a looseness in \eqref{eq:probTime}; however, in Section~\ref{sec:conj}, we conjecture a sparsity-dependent expectation bound that closely matches the empirical results for signals that are sparse in the time domain.
Third, under both models of sparsity, and 
assuming $8\left(\log{2L}+2\right) \ll K$ for signals of practical interest, the expectation bounds on $\rho \left(\vc{a}\right)$ are qualitatively lower than the deterministic bounds. This raises the question of whether the 
deterministic bounds are sharp. We confirm that this is the case in Example~\ref{exm:tight} below.
\begin{example} (Illustrating the signal-dependency of the left-hand side of \ac{CoM} inequalities~\eqref{eq1} and~\eqref{eq2})
In this example, we illustrate that the \ac{CoM} behavior for randomized Toeplitz matrices is indeed signal-dependent. We consider inequality~\eqref{eq1} while a similar analysis can be made for~\eqref{eq2}. We consider two particular $K$-sparse ($K=64$) signals, $\vc{a}_1$ and $\vc{a}_2$ both in $\real^N~(N = 1024)$ where the $K$ non-zero entries of $\vc{a}_1$ have equal values and occur in the first $K$ entries of the vector ($\rho\left(\vc{a}_1\right) = 63.26$), while the $K$ non-zero entries of $\vc{a}_2$ appear in a randomly-selected locations with random signs and values ($\rho\left(\vc{a}_2\right) = 5.47$). Both $\vc{a}_1$ and $\vc{a}_2$ are normalized. For a fixed $M = 512$, we measure each of these signals with $1000$ \ac{i.i.d.} Gaussian $M \times N$ Toeplitz matrices.
Figure~\ref{fig:sig_dep_bnd} depicts the numerically determined rate of occurrence of the event
$\|\vc{y}\|_2^2 - M\|\vc{a}\|_2^2 \geq \epsilon M\|\vc{a}\|_2^2$ over $1000$ trials
versus $\epsilon \in \left(0,1\right)$. For comparison, the derived analytical bound in (\ref{eq1}) (for Toeplitz $X$) as well as the bound in (\ref{eq:conc_unstruct}) (for unstructured $X$) is depicted.
As can be seen the two signals have different concentrations. In particular, $\vc{a}_1$ (Fig.~\ref{fig:a1}) has worse concentration compared to $\vc{a}_2$ (Fig.~\ref{fig:a2}) when measured by Toeplitz matrices. This signal-dependency of the concentration does not happen when these signals are measured by unstructured Gaussian random matrices. Moreover, the derived signal-dependent bound~\eqref{eq1} successfully upper bounds the numerical event rate of occurrence for each signal while the bound (\ref{eq:conc_unstruct}) for unstructured $X$ fails to do so. Also observe that the analytical bound $e^{-\frac{M\epsilon^2}{8\rho\left(\vc{a}_2\right)}}$ in Fig.~\ref{fig:a2} can not bound the numerical event rate of occurrence for $\vc{a}_1$ as depicted in Fig.~\ref{fig:a1}.
\label{exm:sig_dep_bnd} 
\end{example}

\begin{figure}[tb]
\centering
\subfigure[]{
   \includegraphics[width = .45\columnwidth]{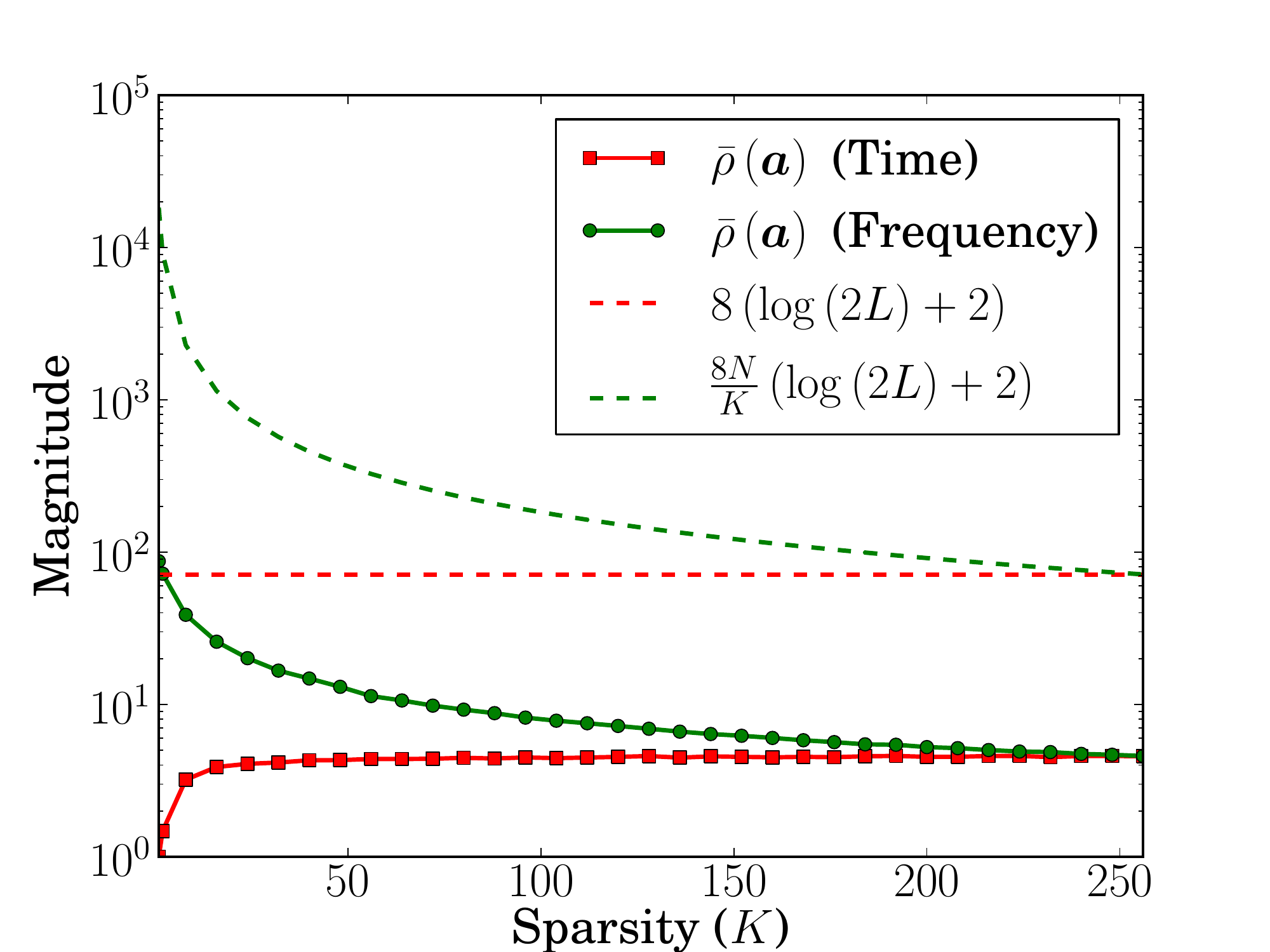}
   \label{fig:basis_256}
 }
\subfigure[]{
   \includegraphics[width = .45\columnwidth]{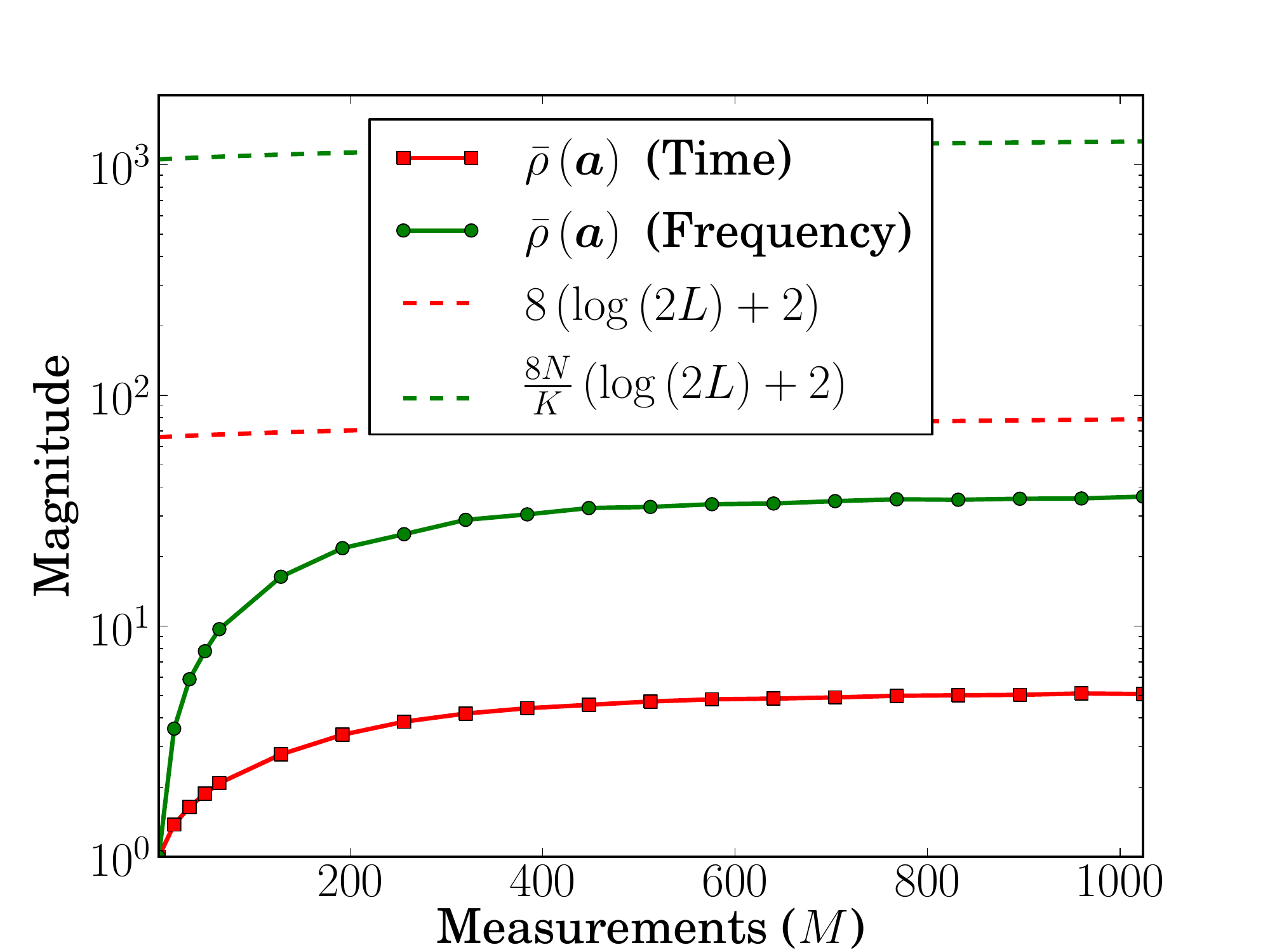}
   \label{fig:dft_N_1024}
 }
\caption{\small\sl Sample mean of $\rho\left(\vc{a}\right)$ in the time and frequency domains versus the expectation bounds \eqref{eq:probTime} and \eqref{eq:probFreq}, where $L = N+M-1$. The sample mean $\bar{\rho}\left(\vc{a}\right)$ is calculated by taking the mean over $1000$ constructed signals $\vc{a}$. A logarithmic scale is used for the vertical axis.
(a) Example~\ref{exm:basis_256}: Varying $K$ with fixed $M = N = 256$.
(b) Example~\ref{exm:dft_1024}: Varying $M$ with fixed $N = 512$ and $K = 32$.}
\end{figure}

\begin{example} (Varying $K$ and comparing the time and frequency domains)
In this experiment, we fix $M$ and $N$. For each value of $K$ and each sparse basis $G=I_N$ and $G=R_{N}$, we construct 1000 random sparse vectors $\vc{q} \in \mathbb{R}^N$ with random support and having $K$ non-zero entries drawn from a Gaussian distribution with mean zero and variance $\frac{1}{K}$. For each vector, we compute $\vc{a}=G\vc{q}$, and we then let $\bar{\rho}\left(\vc{a}\right)$ denote the sample mean of $\rho\left(\vc{a}\right)$ across these 1000 signals. The results, as a function of $K$, are plotted in Fig.~\ref{fig:basis_256}. As anticipated, signals that are sparse in the frequency domain have a larger value of $\bar{\rho}\left(\vc{a}\right)$ than signals that are sparse in the time domain. Moreover, $\bar{\rho}\left(\vc{a}\right)$ decreases with $K$ in the frequency domain but increases with $K$ in the time domain. Overall, the empirical behavior of $\bar{\rho}\left(\vc{a}\right)$ is mostly consistent with the bounds \eqref{eq:probTime} and \eqref{eq:probFreq}, although our constants may be larger than necessary.
\label{exm:basis_256}
\end{example}

\begin{example} (Varying $M$ and comparing the time and frequency domains) This experiment is identical to the one in Example~\ref{exm:basis_256}, except that we vary $M$ while keeping $K$ and $N$ fixed. The results are plotted in Fig.~\ref{fig:dft_N_1024}.
Once again, signals that are sparse in the frequency domain have a larger value of $\bar{\rho}\left(\vc{a}\right)$ than signals that are sparse in the time domain.
Moreover, in both cases $\bar{\rho}\left(\vc{a}\right)$ appears to increase logarithmically with $M$ as predicted by the bounds in~(\ref{eq:probTime}) and (\ref{eq:probFreq}), although our constants may be larger than necessary.
\label{exm:dft_1024}
\end{example}

\begin{example} (Confirming the tightness of the deterministic bounds)
We fix $N$, consider a vector $\vc{a} \in \real^N$ that is $K$-sparse in the time domain, and suppose the $K$ non-zero entries of $\vc{a}$ take equal values and occur in the first $K$ entries of the vector. 
For such a vector one can derive a lower bound on $\rho \left(\vc{a}\right)$ by embedding $P(\vc{a})$ inside a circulant matrix, applying the Cauchy Interlacing Theorem~\cite{hwang2004cauchy} (we describe these steps more fully in Section~\ref{sec:cembed}), and then performing further computations that we omit for the sake of space. 
With these steps, one concludes that for this specific vector $\vc{a} \in \real^N$,
\begin{equation}
\rho\left(\vc{a}\right) \geq K\left(1-\frac{\pi^2}{24}\left(\frac{K^2}{M+K-1}\right)^2\right)^2.
\label{eq:rho_lower_bnd_block_sig}
\end{equation}
When $M \gg K^2$, the right-hand side of~(\ref{eq:rho_lower_bnd_block_sig}) approaches $K$. This confirms that~\eqref{eq:detTime} is sharp for large $M$.
\label{exm:tight}
\end{example}

In the remainder of the paper, the proofs of Theorem~\ref{maintheo1} (Section~\ref{sec:theo1}) and Theorem~\ref{maintheo2} (Section~\ref{sec:theo2}) are presented, followed by additional discussions concerning the relevance of the main results. Our results have important consequences in the analysis of high-dimensional dynamical systems. We expound on this fact by exploring a \ac{CBD} problem in Section~\ref{sec:CBD}.

\section{Proof of Theorem~\ref{maintheo1}}
\label{sec:theo1}
The proofs of the upper and lower bounds of Theorem~\ref{maintheo1} are given separately in Lemmas~\ref{maintheo1_upper_lemma} and \ref{maintheo1_lower_lemma} below.
The proofs utilize Markov's inequality along with a suitable bound on the moment generating function of $\|\vc{y}\|_2^2$, given in~Lemma~\ref{lem:gencovbound}. Observe that for a fixed vector $\vc{a} \in \real^N$ and a random Gaussian Toeplitz matrix $X \in \real^{M \times N}$, the vector $\vc{y} = X\vc{a} \in \real^M$ will be a Gaussian random vector with zero mean and $M \times M$ covariance matrix $P\left(\vc{a}\right)$ given in~(\ref{covmat1}).
\begin{lemma}
If $\vc{y} \in \real^{M}$ is a zero mean Gaussian random vector with covariance matrix $P$, then
\begin{equation}
\E{e^{t \vc{y}^T\vc{y}}} = \frac{1}{\sqrt{\det\left(I_M - 2tP\right)}}
\end{equation}
holds for all $t \in (-\infty, \frac{1}{2\lambda_{\text{max}}(P)})$.
\label{lem:gencovbound}
\end{lemma}

\prf{
\begin{align*}
\E{e^{t \vc{y}^T\vc{y}}}& = \int \frac{1}{(2\pi)^{\frac{M}{2}} \det^{\frac{1}{2}}\left(P\right)} e^{t \vc{y}^T\vc{y}}e^{-\frac{1}{2}\vc{y}^TP^{-1}\vc{y}}d\vc{y}\\
&= \int \frac{1}{(2\pi)^{\frac{M}{2}} \det^{\frac{1}{2}}\left(P\right)} e^{-\frac{1}{2}\vc{y}^T\left(P^{-1}- 2t I_M\right)\vc{y}}d\vc{y}
=\frac{\det^{\frac{1}{2}}\left(\left(P^{-1} - 2t I_M\right)^{-1}\right)} {\det^{\frac{1}{2}}\left(P\right)}\\
&=\frac{1}{\left(\det\left(P^{-1} - 2t I_M\right)\det{P}\right)^{\frac{1}{2}}}
=\frac{1}{\sqrt{\det\left(I_M - 2t P\right)}}.
\end{align*}
}

Observe that as a special case of Lemma~\ref{lem:gencovbound}, if $y \in \real$ is a scalar Gaussian random variable of unit variance, then we obtain the well known result of
$\E{e^{t y^{2}}} = \frac{1}{\sqrt{1 - 2t}}$, for $t \in (-\infty, \frac{1}{2})$.
Based on Lemma~\ref{lem:gencovbound}, we use Chernoff's bounding 
method~\cite{lugosi2004concentration} for computing the upper and lower tail probability bounds. In particular, we are interested in finding bounds for the 
tail probabilities
\begin{subequations}
\begin{equation}
\Prob{\|\vc{y}\|_2^{2} \geq M\|\vc{a}\|_2^2(1+\epsilon)}
\label{uppertailbnd}
\end{equation}
and
\begin{equation}
\ \Prob{\|\vc{y}\|_2^{2} \leq M\|\vc{a}\|_2^2(1-\epsilon)}.
\label{lowertailbnd}
\end{equation}
\label{probtailbnds}
\end{subequations}
Observe that in~(\ref{uppertailbnd}) and~(\ref{lowertailbnd}) concentration behavior is sought around $\E{\norm{\vc{y}}_2^2} = M \norm{\vc{a}}_2^2$.
For a random variable $z$, and all $t>0$,
\begin{equation}
\Prob{z \geq \epsilon} = \Prob{e^{tz} \geq e^{t\epsilon}} \leq \frac{\E{e^{tz}}}{e^{t\epsilon}}
\label{chernoffupper}
\end{equation}
(see, e.g., \cite{lugosi2004concentration}). Applying (\ref{chernoffupper}) to~(\ref{uppertailbnd}), for example, and then applying Lemma~\ref{lem:gencovbound} yields
\begin{equation}
\Prob{\|\vc{y}\|_2^{2} > M\|\vc{a}\|_2^2\left(1+\epsilon\right)} \leq \frac{\E{e^{t\vc{y}^T\vc{y}}}}{e^{M\|\vc{a}\|_2^2\left(1+\epsilon\right)t}} = \left(\det\left(I_M-2tP\right)\right)^{-\frac{1}{2}}
e^{-M\|\vc{a}\|_2^2\left(1+\epsilon\right)t}.
\label{eqn:tbound}
\end{equation}
In~(\ref{eqn:tbound}), $t \in (-\infty,\frac{1}{2(\max_i \lambda_i)})$ is a free variable which---as in a Chernoff bound---can be varied to make the right-hand side as small as possible.
Though not necessarily optimal, we propose to use
$
t = \frac{\epsilon}{2(1+\epsilon)f(\vc{a})\|\vc{a}\|_2^2},
$
where $f$ is a function of $\vc{a}$ that we specify below. We state the upper tail probability bound in Lemma~\ref{maintheo1_upper_lemma} and the lower tail probability bound in Lemma~\ref{maintheo1_lower_lemma}.

\begin{lemma}
Let $\vc{a} \in \real^N$ be fixed, let $P = P\left(\vc{a}\right)$ be as given in~(\ref{covmat1}), and let $\vc{y} \in \real^{M}$ be a zero mean Gaussian random vector with covariance matrix $P$. Then, for any $\epsilon \in (0,1)$,
\begin{equation}
\Prob{\|\vc{y}\|_2^{2} \geq M\|\vc{a}\|_2^2\left(1+\epsilon\right)} \leq e^{-\frac{\epsilon^{2}M}{8\rho\left(\vc{a}\right)}}.
\label{equpper1}
\end{equation}
\label{maintheo1_upper_lemma}
\end{lemma}

\prf
{
Choosing $t$ as
\begin{equation*}
t = \frac{\epsilon}{2\left(1+\epsilon\right)\rho\left(\vc{a}\right)\|\vc{a}\|_2^2} 
\end{equation*}
and noting that $t \in (-\infty, \frac{1}{2\max_i\lambda_i})$, the right-hand side of (\ref{eqn:tbound}) can be written as
\begin{equation}
\left(\left(\det\left(I_M-\frac{\epsilon}{(1+\epsilon)}\frac{P}{\rho\left(\vc{a}\right)
\|\vc{a}\|_2^2}\right)\right)^{-\frac{1}{M}}e^{-\frac{\epsilon}{\rho\left(\vc{a}\right)}}\right)^{\frac{M}{2}}.
\label{rhsupper}
\end{equation}
This expression can be simplified. Note that
\begin{align*}
\det\left(I_M-\frac{\epsilon}{(1+\epsilon)}\frac{P}{\rho\left(\vc{a}\right)\|\vc{a}\|_2^2}\right) & =\prod_{i=1}^{M} \left(1 - \frac{\epsilon}{(1+\epsilon)}\frac{\lambda_{i}}{\rho\left(\vc{a}\right)\|\vc{a}\|_2^2}\right)\\
&= e^{\sum_{i=1}^{M}\log\left(1-\frac{\epsilon}{(1+\epsilon)}\frac{\lambda_{i}}{\rho\left(\vc{a}\right)
\|\vc{a}\|_2^2}\right)}.
\end{align*}
Using the facts that $\log \left(1-c_1c_2\right) \geq c_2 \log \left(1-c_1\right)$ for any $c_1,c_2 \in [0,1]$ and $\Tr{P}= M\|\vc{a}\|_2^2$, we have
\begin{align}
e^{\sum_{i=1}^{M}\log\left(1-\frac{\epsilon}{(1+\epsilon)}\frac{\lambda_{i}}{\rho\left(\vc{a}\right)
\|\vc{a}\|_2^2}\right)}
& \geq
e^{\sum_{i=1}^{M}\frac{\lambda_{i}}{\rho\left(\vc{a}\right)
\|\vc{a}\|_2^2}\log\left(1-\frac{\epsilon}{1+\epsilon}\right)} \notag \\
& =
e^{\frac{\Tr{P}}{\rho\left(\vc{a}\right)\|\vc{a}\|_2^2}\log \left(\frac{1}{1+\epsilon}\right)}
=
e^{\frac{M}{\rho\left(\vc{a}\right)}\log \left(\frac{1}{1+\epsilon}\right)}
=\left(\frac{1}{1+\epsilon}\right)^{\frac{M}{\rho\left(\vc{a}\right)}}.
\label{logupper}
\end{align}
Combining~(\ref{eqn:tbound}),~(\ref{rhsupper}), and~(\ref{logupper}) gives us
\[
\Prob{\|\vc{y}\|_2^{2} > M\|\vc{a}\|_2^2\left(1+\epsilon\right)} \leq  \left(\left(\frac{1}{1+\epsilon}\right)^{-\frac{1}{\rho\left(\vc{a}\right)}} e^{-\frac{\epsilon}{\rho\left(\vc{a}\right)}}\right)^{\frac{M}{2}} =\left(\left(1+\epsilon \right)e^{-\epsilon}\right)^{\frac{M}{2\rho\left(\vc{a}\right)}}.
\]
The final bound comes by noting that $(1+\epsilon)e^{-\epsilon} \leq e^{-\epsilon^2/4}$.
}


\begin{lemma} Using the same assumptions as in Lemma~\ref{maintheo1_upper_lemma}, for any $\epsilon \in (0,1)$,
\[
\Prob{\|\vc{y}\|_2^{2} \leq M\|\vc{a}\|_2^2\left(1-\epsilon\right)} \leq e^{-\frac{\epsilon^{2}M}{8\mu\left(\vc{a}\right)}}.
\]
\label{maintheo1_lower_lemma}
\end{lemma}

\prf
{
Applying Markov's inequality to~(\ref{lowertailbnd}), we obtain
\begin{equation}
\Prob{\|\vc{y}\|_2^{2} \leq M\|\vc{a}\|_2^2(1-\epsilon)} = \Prob{-\|\vc{y}\|_2^{2} \geq -M\|\vc{a}\|_2^2(1-\epsilon)}
\leq \frac{\E{e^{-t\vc{y}^T\vc{y}}}}{e^{-M\|\vc{a}\|_2^2(1-\epsilon)t}}.
\label{lowertailprob}
\end{equation}
Using Lemma~\ref{lem:gencovbound}, this implies that
\begin{equation}
\label{eqn:tboundlow}
\Prob{\|\vc{y}\|_2^{2} \leq M\|\vc{a}\|_2^2(1-\epsilon)} \leq \\
\left(\left(\det\left(I_M+2tP\right)\right)^{-\frac{1}{M}}e^{2\|\vc{a}\|_2^2(1-\epsilon)t}\right)^{\frac{M}{2}}.
\end{equation}
In this case, we choose
\begin{equation*}
t = \frac{\epsilon}{2(1+\epsilon)\mu(\vc{a})\|\vc{a}\|_2^2},
\end{equation*}
and note that $t > 0$. Plugging $t$ into~(\ref{eqn:tboundlow}) and following similar steps as for the upper tail bound, we get
\begin{equation}
\det(I_M+2tP) 
=\prod_{i=1}^{M}\left(1 + \frac{\epsilon}{(1+\epsilon)}\frac{\lambda_{i}}{\mu(\vc{a})\|\vc{a}\|_2^2}\right)
= e^{\sum_{i=1}^{M}\log\left(1 +\frac{\epsilon}{\left(1+\epsilon\right)}\frac{\lambda_{i}}{\mu(\vc{a})\|\vc{a}\|_2^2}\right)}. \label{eqn:detbndpos}
\end{equation}
Since $\log\left(1+c\right) \geq c-\frac{c^2}{2}$ for $c>0$,
\begin{align}
\sum_{i=1}^{M}
\log\left(1+\frac{\epsilon}{\left(1+\epsilon\right)}\frac{\lambda_{i}}{\mu\left(\vc{a}\right)
\|\vc{a}\|_2^2}\right)
&\geq \sum_{i=1}^{M} \left(\frac{\epsilon}{\left(1+\epsilon\right)}\frac{\lambda_{i}}{\mu\left(\vc{a}\right)\|\vc{a}\|_2^2}-
\frac{1}{2}\left(\frac{\epsilon}{\left(1+\epsilon\right)}\frac{\lambda_{i}}{\mu\left(\vc{a}\right)
\|\vc{a}\|_2^2}\right)^2\right) \notag \\
&= \frac{\epsilon}{\left(1+\epsilon\right)}\frac{\sum_{i=1}^{M}\lambda_{i}}{\mu\left(\vc{a}\right)\|\vc{a}\|_2^2}  - \frac{1}{2}\left(\frac{\epsilon}{\left(1+\epsilon\right)\mu\left(\vc{a}\right)\|\vc{a}\|_2^2}\right)^2 \sum_{i=1}^{M} \lambda_{i}^2 \notag \\
&= \frac{\epsilon}{\left(1+\epsilon\right)}\frac{M}{\mu\left(\vc{a}\right)}-
\frac{1}{2}\left(\frac{\epsilon}{1+\epsilon}\right)^2\frac{M}{\mu\left(\vc{a}\right)}
= \frac{M}{\mu\left(\vc{a}\right)} \left(\frac{\epsilon^2+2\epsilon}{2\left(1+\epsilon\right)^2}\right).
\label{eq:aux_lower1}
\end{align}
Combining (\ref{eqn:detbndpos}) and (\ref{eq:aux_lower1}) gives the bound
\begin{align}
\det\left(I_M+2tP\right)
& \geq
e^{\frac{M}{\mu\left(\vc{a}\right)}\left(\frac{\epsilon^2+2\epsilon}{2\left(1+\epsilon\right)^2}\right)}
=\left(e^{\frac{\epsilon^2+2\epsilon}{2\left(1+\epsilon\right)^2}}\right)^{\frac{M}{\mu\left(\vc{a}\right)}}. \label{loglower}
\end{align}
By substituting~(\ref{loglower}) into (\ref{eqn:tboundlow}), we obtain
\begin{align*}
\Prob{\|\vc{y}\|_2^{2} \leq M\|\vc{a}\|_2^2\left(1-\epsilon\right)} &\leq  \left(e^{\frac{-\epsilon^2-2\epsilon}{2\left(1+\epsilon\right)^2}} e^{\frac{\epsilon\left(1-\epsilon\right)}{1+\epsilon}}\right)^{\frac{M}{2\mu\left(\vc{a}\right)}}
=\left(e^{\frac{-2\epsilon^3-\epsilon^2}{2\left(1+\epsilon\right)^2}}\right)^{\frac{M}{2\mu\left(\vc{a}\right)}}.
\end{align*}
The final bound comes by noting that $e^{\frac{-2\epsilon^3-\epsilon^2}{2\left(1+\epsilon\right)^2}} \leq e^{-\epsilon^2/4}$.
}

\section{Proof and Discussion of Theorem~\ref{maintheo2}}
\label{sec:theo2}

\subsection{Proof of Theorem~\ref{maintheo2}}

%

\subsubsection{Circulant Embedding}
\label{sec:cembed}

The covariance matrix $P\left(\vc{a}\right)$ described in~(\ref{covmat1}) is an $M \times M$ symmetric Toeplitz matrix which can be decomposed as
$
P\left(\vc{a}\right) = A^TA,
$
where $A$ is an $(N+M-1) \times M$ Toeplitz matrix (as shown in Fig.~\ref{fig:toep_circ})
and $A^T$ is the transpose of $A$. 
\begin{figure}[tb]
\begin{minipage}[b]{0.45\linewidth}
\centering
\begin{equation*}
A =
\left[
\begin{array}{ccccc}
    a_1 & 0 & \hdots & & 0 \\
    a_2 & \ddots & & \left(0\right)&  \\
    \vdots & \ddots & a_1& & \vdots\\
    a_N &  & a_2 & \ddots& 0\\
    0 & \ddots & \vdots & \ddots & a_1 \\
    \vdots& & a_N & & a_2\\
     & \left(0\right)& & \ddots & \vdots\\
     0 & & \hdots& 0 & a_N
\end{array}
\right]
\end{equation*}
\end{minipage}
\hspace{0.05cm}
\begin{minipage}[b]{0.45\linewidth}
\centering
\begin{equation*}
A_c =
\left[
\begin{array}{cccccccc}
    a_1 & 0 & \hdots & & 0 & a_N &\hdots &a_2\\
    a_2 & \ddots & \ddots& & & \ddots& \ddots &\vdots \\
    \vdots & \ddots & a_1& &\left(0\right) & & \ddots& a_N\\
    a_N &  & a_2 & \ddots& & & & 0\\
    0 & \ddots & \vdots & \ddots & a_1 & & &\\
    \vdots& & a_N & & a_2&\ddots & \ddots& \vdots\\
     & \left(0\right)& & \ddots & \vdots& \ddots & \ddots & 0\\
     0 & & \hdots& 0 & a_N&\hdots & a_2 & a_1
\end{array}
\right]
\end{equation*}
\end{minipage}
\caption{\small\sl Toeplitz matrix $A \in \real^{L\times M}$ and its circulant counterpart $A_c \in \real^{L \times L}$ where $L = N + M -1$.}
\label{fig:toep_circ}
\end{figure}
In order to derive an upper bound on the maximum eigenvalue of
$P\left(\vc{a}\right)$, we embed the matrix $A$ inside its $(N+M-1) \times (N+M-1)$ circulant counterpart $A_c$ where each column of $A_c$ is a cyclic downward shifted version of the previous column. Thus, $A_c$ is uniquely determined by its first column, which we denote by
\begin{equation*}
\widetilde{\vc{a}} = [\underbrace{a_1 \ \cdots \ a_N}_{\vc{a}^T} \
\underbrace{0 \ \cdots \ 0}_{\left(M-1\right) \ \text{zeros}}]^T \in \real^{L},
\end{equation*}
where $L = N+M-1$. Observe that the circulant matrix $A_c \in \real^{L\times L}$ contains the Toeplitz matrix $A \in \real^{L \times M}$ in its first $M$ columns.
Because of this embedding, the Cauchy Interlacing Theorem~\cite{hwang2004cauchy} implies that $\max_{i\leq M} \lambda_i(A^TA) \leq \max_{i\leq L} \lambda_i(A_c^T A_c)$.
Therefore, we have
\begin{equation}
\rho \left(\vc{a}\right) = \frac{\max_i \lambda_i(P\left(\vc{a}\right))}{\|\vc{a}\|_2^2} = \frac{\max_i \lambda_i(A^TA)}{\|\vc{a}\|_2^2} \leq \frac{\max_i \lambda_i(A_c^T A_c)}{\|\vc{a}\|_2^2} = \frac{\max_{i} |\lambda_i(A_c^T)|^2}{\|\vc{a}\|_2^2} =: \rho_c \left(\vc{a}\right).
\label{eq:rho_upper_bnd_1}
\end{equation}
Thus, an upper bound for $\rho\left(\vc{a}\right)$ can be achieved 
by bounding the maximum absolute eigenvalue of $A_c^T$.
Since $A_c^T$ is circulant, its eigenvalues are given by the un-normalized length-$L$ \ac{DFT} of the first row of $A_c^T$ (the first column of $A_c$).
Specifically, for $i = 1, 2, \dots, L$,
\begin{equation}
\lambda_i(A_c^T) = \sum_{k=1}^{L} \widetilde{a}_k e^{-\frac{2\pi j}{L}(i-1)(k-1)} = \sum_{k=1}^{N} a_k e^{-\frac{2\pi j}{L}(i-1)(k-1)}.
\label{eq:eig_summation}
\end{equation}
Recall that $F_L \in \mathbb{C}^{L \times L}$ is the Fourier orthobasis with entries $F_L(\ell, m) = \frac{1}{\sqrt{L}}w^{(\ell-1)(m-1)}$
where
$w = e^{-\frac{2\pi j}{L}}$, and let $F_L^{i\to} \in \mathbb{C}^L$ be the $i$-th row of $F_L$. Using matrix-vector notation, (\ref{eq:eig_summation}) can be written as
\begin{equation}
\lambda_i (A_c^T) = \sqrt{L}F_L^{i\to}\widetilde{\vc{a}} = \sqrt{L}F^{i\to}_{1:N}\vc{a} = \sqrt{L}F^{i\to}_{1:N}G\vc{q} = \sqrt{L}F^{i\to}_{1:N}G_S\vc{q}_S,
\label{eq:eig_product_format}
\end{equation}
where $F^{i\to}_{1:N} \in \mathbb{C}^N$ is a row vector containing the first $N$ entries of $F_L^{i\to}$, $\vc{q}_S \in \real^K$ is the part of $\vc{q} \in \real^N$ restricted to the support $S$ (the location of the non-zero entries of $\vc{q}$) with cardinality $\left|{S}\right| = K$, and $G_S \in \real^{N \times K}$ contains the columns of $G \in \real^{N \times N}$ indexed by the support $S$.


\subsubsection{Deterministic Bound}

We can bound $\rho \left(\vc{a}\right)$ over all sparse $\vc{a}$ using the Cauchy-Schwarz inequality. From~(\ref{eq:eig_product_format}), it follows for any $i \in \{1, 2, \dots, L\}$ that
\begin{equation}
|\lambda_i (A_c^T)| = |\sqrt{L}F^{i\to}_{1:N}G_S\vc{q}_S| \leq \sqrt{L}\|F^{i\to}_{1:N}G_S\|_2 \|\vc{q}_S\|_2 = \sqrt{L}\|F^{i\to}_{1:N}G_S\|_2 \|\vc{a}\|_2.
\label{eq:cauchy_1}
\end{equation}
By combining Definition~\ref{def:nu1}, (\ref{eq:rho_upper_bnd_1}), and~(\ref{eq:cauchy_1}), we arrive at the deterministic bound~\eqref{eq:consbnd}.
%
This bound appears to be highly pessimistic for {\em most} sparse vectors $\vc{a}$. In other words, although in Example~\ref{exm:tight} we illustrate that for a specific signal $\vc{a}$, the deterministic bound~\eqref{eq:consbnd} is tight when $M\gg K$, we observe that for many other classes of sparse signals $\vc{a}$, the bound is pessimistic. In particular, if a random model is imposed on the non-zero entries of $\vc{a}$, an upper bound on the typical value of $\rho\left(\vc{a}\right)$ derived in (\ref{eq:probTime}) scales logarithmically in the ambient dimension $L$ which is qualitatively smaller than $K$. We show this analysis in the proof of Theorem~\ref{maintheo2}. In order to make this proof self-contained, we first list some results that we will draw from.

%
\subsubsection{Supporting Results}
We utilize the following propositions.

\begin{lemma}{\cite{rauhut2010compressive}}
Let $z$ be any random variable. Then 
\begin{equation}
\E{|z|} = \int_0^{\infty} \Prob{|z| \geq x}dx.
\end{equation}
\label{lem:absolute_moment}
\end{lemma}

\begin{lemma}
Let $z_1$ and $z_2$ be positive random variables. Then for any $U$,
\begin{equation}
\Prob{z_1 + z_2 \geq U} \leq \Prob{z_1 \geq \frac{U}{2}} + \Prob{z_2 \geq \frac{U}{2}},
\label{coro:prob_1_sum}
\end{equation}
and for any $U_1$ and $U_2$,
\begin{equation}
\Prob{\frac{z_1}{z_2}\geq \frac{U_1}{U_2}} \leq \Prob{z_1 \geq U_1} + \Prob{z_2 \leq U_2}.
\label{coro:prob_1_product}
\end{equation}
\label{coro:prob_1}
\end{lemma}

\prf{See Appendix A.}

\begin{proposition}{\cite{achlioptas2003database}~({\em Concentration Inequality for Sums of Squared Gaussian Random Variables})}
Let $\vc{q} \in \mathbb{R}^N$ be a random $K$-sparse vector whose $K$ non-zero entries (on an arbitrary support $S$) are \ac{i.i.d.} random variables drawn from a Gaussian distribution with $\mathcal{N} (0, \sigma^2)$. Then for any $\epsilon > 0$,
\begin{equation*}
\Prob{\|\vc{q}\|_2^2 \leq K\sigma^2\left(1-\epsilon\right)} \leq e^{-\frac{K\epsilon^2}{4}}.
\end{equation*}
\label{prop:com_sum_square}
\end{proposition}

\begin{proposition}{~({\em Hoeffding's Inequality for Complex-Valued Gaussian Sums})}
Let $\vc{b} \in \mathbb{C}^N$ be fixed, and let $\vc{\epsilon} \in \mathbb{R}^N$ be a random vector whose $N$ entries are \ac{i.i.d.} random variables drawn from a Gaussian distribution with $\mathcal{N} (0, \sigma^2)$. Then, for any $u > 0$,
\begin{equation*}
\Prob{\left|\sum_{i = 1}^{N}\epsilon_ib_i\right|\geq u} \leq 2e^{-\frac{u^2}{4\sigma^2\|\vc{b}\|_2^2}}.
\end{equation*}
\label{propos:hoeff_complex_Gauss}
\end{proposition}

\prf{See Appendix B.}

In order prove Theorem~\ref{maintheo2}, we also require a tail probability bound for the eigenvalues of $A^T_c$.
\begin{proposition}
Let $\vc{q} \in \mathbb{R}^N$ be a random $K$-sparse vector whose $K$ non-zero entries (on an arbitrary support $S$) are i.i.d.\ random variables drawn from a Gaussian distribution with $\mathcal{N} (0, \frac{1}{K})$. Let $\vc{a} = G\vc{q}$ where $G \in \real^{N \times N}$ is an orthobasis, and let $A_c$ be an $L \times L$ circulant matrix, where the first $N$ entries of the first column of $A_c$ are given by $\vc{a}$. Then for any $u > 0$, and for $i = 1, 2, \dots, L$,
\begin{equation}
\Prob{|\lambda_i(A_c)| \geq u} \leq 2e^{-\frac{u^2K}{4L\nu_K^2\left(G\right)}}.
\label{eq:circ_tail_bnd_generic}
\end{equation}
\label{propos:circ_tail_bnd_generic}
\end{proposition}

\prf
{Define the row vector $\vc{b} = \sqrt{L}F^{i\to}_{1:N}G_S \in \mathbb{C}^K$. From (\ref{eq:eig_product_format}) and the Cauchy-Schwarz inequality, it follows that $|\lambda_i(A_c)| =  |\lambda_i (A_c^T)| = |\sqrt{L}F^{i\to}_{1:N}G_S\vc{q}_S| = | \sum_{i = 1}^K \epsilon_i b_i |$, where $\epsilon_i = (\vc{q}_S)_i$. From Definition~\ref{def:nu1}, we have $\|\vc{b}\|_2 \leq \sqrt{L}\nu_K\left(G\right)$. The tail probability bound~\eqref{eq:circ_tail_bnd_generic} follows from applying Proposition~\ref{propos:hoeff_complex_Gauss}.
}


\subsubsection{Completing the Proof of Theorem~\ref{maintheo2}}

From (\ref{eq:rho_upper_bnd_1}), we have
\begin{align*}
\E{\rho\left(\vc{a}\right)} \leq \E{\rho_c\left(\vc{a}\right)} = \E{\frac{\max_{i} |\lambda_i(A_c^T)|^2}{\|\vc{a}\|_2^2}}
&= \int_0^{\infty} \Prob{\frac{\max_{i} |\lambda_i(A_c^T)|^2}{\|\vc{a}\|_2^2} \geq x}dx \\
&= \int_0^{L\nu_G^2} \Prob{\frac{\max_{i} |\lambda_i(A_c^T)|^2}{\|\vc{a}\|_2^2} \geq x}dx,
\end{align*}
where the last equality comes from the deterministic upper bound $|\lambda_i (A_c^T)| \leq \sqrt{L}\|F^{i\to}_{1:N}G_S\|_2 \|\vc{a}\|_2 \leq \sqrt{L}\nu_K\left(G\right) \|\vc{a}\|_2$.
Using a union bound, for any $t>0$ we have
\begin{align}
\int_0^{L\nu_K^2\left(G\right)} \Prob{\frac{\max_{i} |\lambda_i(A_c^T)|^2}{\|\vc{a}\|_2^2} \geq x}dx
&= \int_0^{L\nu_K^2\left(G\right)} \Prob{\frac{\max_{i} |\lambda_i(A_c^T)|^2}{\|\vc{a}\|_2^2} \geq \frac{tx}{t}}dx \notag\\
&\leq \int_0^{L\nu_K^2\left(G\right)} \Prob{\max_{i} |\lambda_i(A_c^T)|^2\geq tx}dx \notag\\
& \ \ \ \ \ + \int_0^{L\nu_K^2\left(G\right)} \Prob{\|\vc{a}\|_2^2 \leq t}dx.
\label{eq:first_term1}
\end{align}
The first term in the right hand side of~(\ref{eq:first_term1}) can be bounded as follows. For every $\delta \geq 0$, by partitioning the range of integration~\cite{rauhut2010compressive,ledoux1991probability}, we obtain
\begin{align*}
\int_0^{L\nu_K^2\left(G\right)} \Prob{\max_{i} |\lambda_i(A_c^T)|^2\geq tx}dx
&\leq \int_0^{\infty} \Prob{\max_{i} |\lambda_i(A_c^T)|^2\geq tx}dx\\
&\leq \delta + \int_{\delta}^{\infty} \Prob{\max_i|\lambda_i(A_c)|^2 \geq tx}dx\\
& \leq \delta + \int_{\delta}^{\infty} \sum_{i=1}^L \Prob{|\lambda_i(A_c)|^2 \geq tx}dx\\
& \le \delta + \int_{\delta}^{\infty} \sum_{i=1}^L 2e^{-\frac{Ktx}{4L\nu_K^2\left(G\right)}} dx\\
& = \delta + 2L\int_{\delta}^{\infty} e^{-\frac{Ktx}{4L\nu_K^2\left(G\right)}}dx\\
& = \delta + \frac{8L^2\nu_K^2\left(G\right)}{Kt}e^{-\frac{Kt\delta}{4L\nu_K^2\left(G\right)}},
\end{align*}
where we used Proposition~\ref{propos:circ_tail_bnd_generic} in the last inequality. The second term in~(\ref{eq:first_term1}) can be bounded using the concentration inequality of Proposition~\ref{prop:com_sum_square}. We have for $0 < t \leq 1$,
$
\Prob{\|\vc{a}\|_2^2 \leq t} \leq e^{-\frac{K\left(1-t\right)^2}{4}}.
$
Putting together the bounds for the two terms of inequality~(\ref{eq:first_term1}), we have
\begin{equation}
\E{\rho\left(\vc{a}\right)} \leq \E{\rho_c\left(\vc{a}\right)}
\leq \delta + \frac{8L^2\nu_K^2\left(G\right)}{Kt}e^{-\frac{Kt\delta}{4L\nu_K^2\left(G\right)}} + L\nu_K^2\left(G\right)e^{-\frac{K\left(1-t\right)^2}{4}}.
\label{eq:tobe min_upper_bnd}
\end{equation}
Now we pick $\delta$ to minimize the upper bound in~(\ref{eq:tobe min_upper_bnd}). Using the minimizer $\delta^{\star} = \frac{4L\nu_K^2\left(G\right)\log{2L}}{Kt}$ yields
\begin{equation}
\E{\rho\left(\vc{a}\right)} \leq \E{\rho_c\left(\vc{a}\right)} \leq \frac{4L\nu_K^2\left(G\right)}{Kt}\left(\log{2L}+1+\frac{Kt}{4}e^{-\frac{K\left(1-t\right)^2}{4}}\right).
\label{eq:choose_t}
\end{equation}
Let $g\left(K,t\right) := \frac{Kt}{4}e^{-\frac{K\left(1-t\right)^2}{4}}$. It is trivial to show that $g\left(K,0.5\right) \leq 1$ for all $K$ (for $t = 0.5, \ \max_K g\left(K,0.5\right) = \frac{2}{e}$). Therefore, $\E{\rho\left(\vc{a}\right)} \leq \frac{8L\nu_K^2\left(G\right)}{K}\left(\log{2L}+2\right)$, which completes the proof.
\hfill $\blacksquare$
%

\subsection{Discussion}
\label{sec:conj}

\begin{remark}
In Theorem~\ref{maintheo2}, we find an upper bound on $\E{\rho\left(\vc{a}\right)}$ by finding an upper bound on $\E{\rho_c\left(\vc{a}\right)}$ and using the fact that for all vectors $\vc{a}$, we have $\rho\left(\vc{a}\right) \leq \rho_c\left(\vc{a}\right)$. However, we should note that this inequality gets tighter as $M$ (the number of columns of $A$) increases. For small $M$ the interlacing technique results in a looser bound.
\end{remark}

\begin{remark}
By taking $G = I_N$ and noting that $\nu_K\left(I_N\right) = \sqrt{\frac{K}{L}}$, \eqref{eq:choose_t} leads to an upper bound on $\E {\rho_c\left(\vc{a}\right)}$ when the signal $\vc{a}$ is $K$-sparse in the time domain (specifically, $\E{\rho_c\left(\vc{a}\right)} \leq 8\left(\log{2L}+2\right)$). Although this bound scales logarithmically in the ambient dimension $L$, it does not show a dependency on the sparsity level $K$ of the vector $\vc{a}$. Over multiple simulations where we have computed the sample mean $\bar{\rho}_c\left(\vc{a}\right)$, we have observed a linear behavior of the quantity $\frac{K}{\bar{\rho}_c\left(\vc{a}\right)}$ as $K$ increases, and this leads us to the conjecture below. Although at this point we are not able to prove the conjecture, the proposed bound matches closely with empirical data. 

\begin{conjecture}
Fix $N$ and $M$. Let $\vc{a} \in \mathbb{R}^N$ be a random $K$-sparse vector whose $K$ non-zero entries (on an arbitrary support $S$) are i.i.d.\ random variables drawn from a Gaussian distribution with $\mathcal{N} (0, \frac{1}{K})$. Then
\begin{equation*}
\E {\rho_c\left(\vc{a}\right)} \sim \frac{K}{c_1K + c_2},
\end{equation*}
where
$c_1 = \frac{1}{c\log{L}}$ for some constant $c$, and $c_2 = 1-c_1$.
\label{conj:expected}
\end{conjecture}

The conjecture follows from our empirical observation that $\frac{K}{\bar{\rho}_c\left(\vc{a}\right)} \sim c_1K+c_2$ for some constants $c_1$ and $c_2$, the fact that $\rho_c\left(\vc{a}\right) = 1$ for $K = 1$, and the observation that $\bar{\rho}_c\left(\vc{a}\right) \sim c\log{L}$ when $K = N$ for large $N$. In the following examples, we illustrate these points and show how the conjectured bound can sharply approximate the empirical mean of $\rho_c\left(\vc{a}\right)$.

\begin{example}
In this experiment, we fix $M = 256$ and take $G = I_N$. For each value of $N$, we construct 1000 random non-sparse vectors $\vc{a} \in \mathbb{R}^N$ whose $N$ entries are drawn from a Gaussian distribution with mean zero and variance $\frac{1}{N}$. We let $\bar{\rho}_c\left(\vc{a}\right)$ denote the sample mean of $\rho_c\left(\vc{a}\right)$ across these 1000 signals. The results, as a function of $N$, are plotted in Fig.~\ref{fig:conjecture_rho_c}. Also plotted is the function $f\left(L\right) = \log\left(L\right)$ where $L = N+M-1$; this closely approximates the empirical data.
\label{exm:conjecture_bnd_rho_c}
\end{example}

\begin{figure}[tb]
\centering
\includegraphics[width = .5\columnwidth]{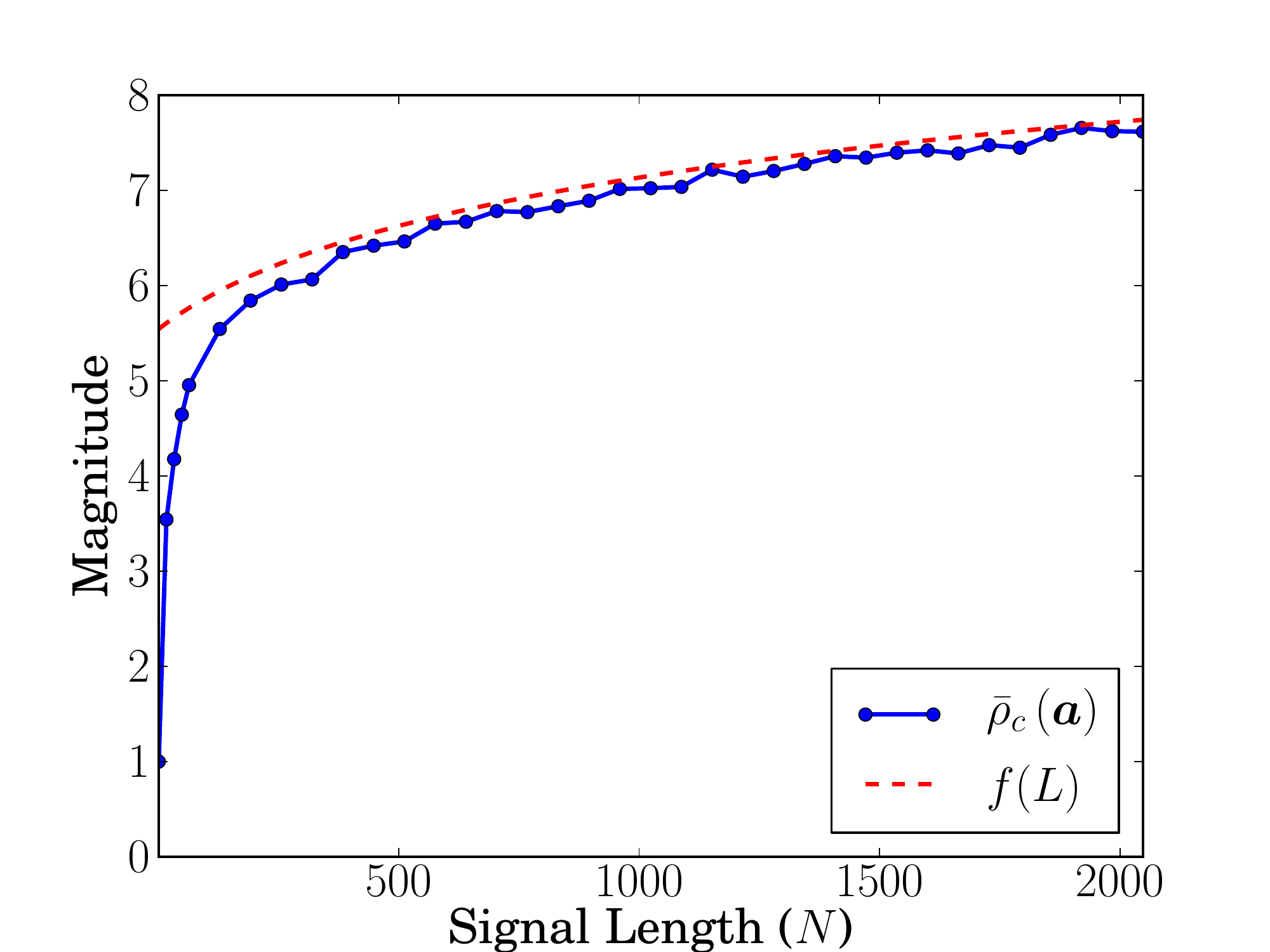}
\caption{\small\sl Empirical results corresponding to Example~\ref{exm:conjecture_bnd_rho_c}: Sample mean of $\rho_c\left(\vc{a}\right)$ in the time domain for full vectors $\vc{a} \in \real^N$ where $M = 256$ is fixed. Also plotted is $f\left(L\right) = \log\left(L\right)$, where $L = N+M-1$.}
\label{fig:conjecture_rho_c}
\end{figure}

\begin{example}
In this experiment, we fix $N=1024$. For each value of $K$, we construct 1000 random sparse vectors $\vc{a} \in \mathbb{R}^N$ with random support and having $K$ non-zero entries drawn from a Gaussian distribution with mean zero and variance $\frac{1}{K}$. We let $\bar{\rho}_c\left(\vc{a}\right)$ denote the sample mean of $\rho_c\left(\vc{a}\right)$ across these 1000 signals. The results, as a function of $K$ for two fixed values $M = 1$ and $M = 1024$, are plotted in Fig.~\ref{fig:dft_K_1024}.
\label{exm:dft_sparsity}
\end{example}

\begin{figure}[tb]
\centering
\subfigure[]{
   \includegraphics[width = .45\columnwidth]{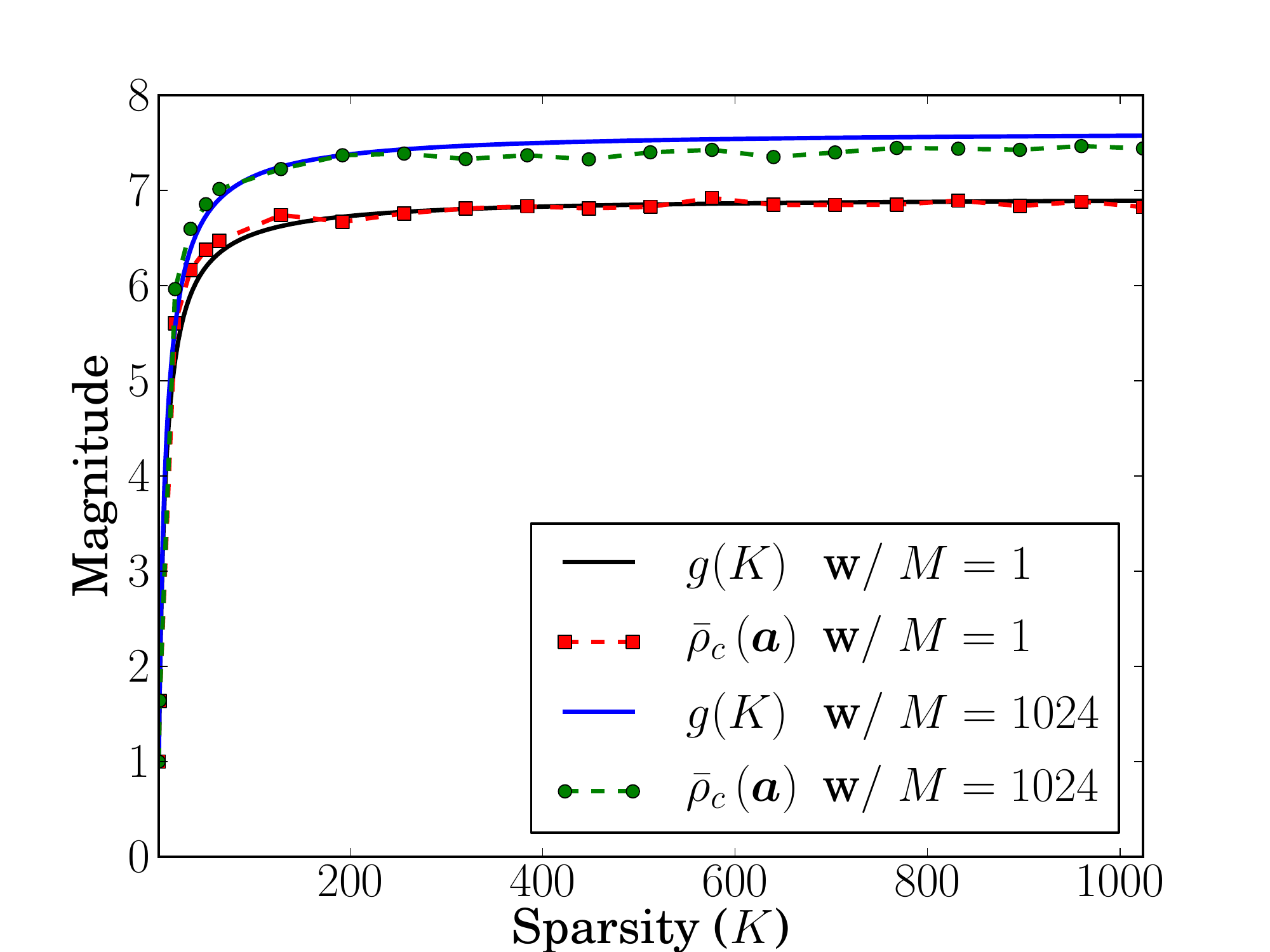}
 }
\subfigure[]{
   \includegraphics[width = .45\columnwidth]{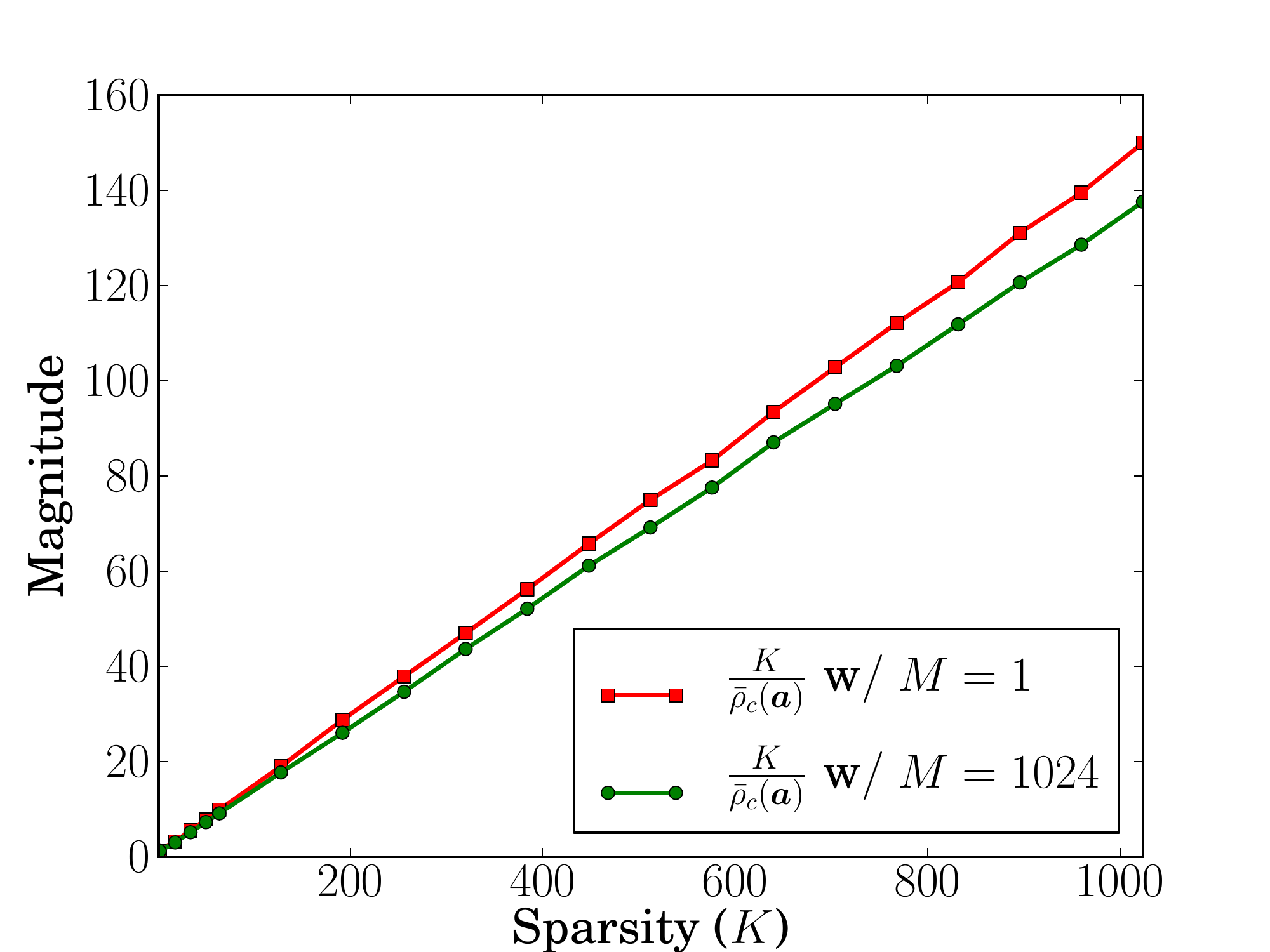}
 }
\caption{\small\sl Empirical results corresponding to Example~\ref{exm:dft_sparsity}. (a) Simulation results vs. the conjectured bound $g(K) = \frac{K}{c_1K+c_2}$ with $c = 1$. (b) Linearity of $\frac{K}{\bar{\rho}_c\left(\vc{a}\right)}$.}
\label{fig:dft_K_1024}
\end{figure}

\end{remark}
\begin{remark}
As a final note in this section, the result of Theorem~\ref{maintheo2} can be easily extended to the case when $G \in \mathbb{C}^{N \times N}$ is a complex orthobasis and $\vc{q}$ and $\vc{a}$ are complex vectors. The bounds can be derived in a similar way and we do not state them for the sake of saving space.
\end{remark}

\subsection{A Quadratic \ac{RIP} Bound and Non-uniform Recovery}
\label{sec:conj}

An approach identical to the one taken by Baraniuk et al.~\cite{baraniuk2008simple} can be used to establish the \ac{RIP} for Toeplitz matrices $X$ based on the \ac{CoM} inequalities given in Theorem \ref{maintheo1}.
As mentioned in Section~\ref{sec:main_result}, the bounds of the \ac{CoM} inequalities for Toeplitz matrices are looser by a factor of $2\rho\left(\vc{a}\right)$ or $2\mu\left(\vc{a}\right)$ as compared to the ones for unstructured $X$.
Since $\rho\left(\vc{a}\right)$ is bounded by $K$ for all $K$-sparse signals in the time domain (the deterministic bound), with straightforward calculations a quadratic estimate of the number of measurements in terms of sparsity ($M \sim K^2$) can be achieved for Toeplitz matrices. 
As mentioned earlier, on the other hand, there exists an extremely non-uniform distribution of $\rho\left(\vc{a}\right)$ over the set of all $K$-sparse signals $\vc{a}$, for as Theorem~\ref{maintheo2} states, if a random model is imposed on $\vc{a}$, an upper bound on the typical value of $\rho\left(\vc{a}\right)$ scales logarithmically in the ambient dimension $L$. This suggests that for most $K$-sparse signals $\vc{a}$ the value of $\rho\left(\vc{a}\right)$ is much smaller than $K$ (observe that $8\left(\log 2L +2\right) \ll K$ for many signals of practical interest). 
Only for a very small set of signals does the value of $\rho\left(\vc{a}\right)$ approach the deterministic bound of $K$. One can show, for example, that for any $K$-sparse signal whose $K$ non-zero entries are all the same, we have $\rho\left(\vc{a}\right) \leq \rho_c\left(\vc{a}\right) = K$ (Example~\ref{exm:tight}). 
This non-uniformity of $\rho\left(\vc{a}\right)$ over the set of sparse signals may be useful for proving a non-uniform recovery bound or for strengthening the \ac{RIP} result; our work on these fronts remains in progress. Using different techniques than pursued in the present paper (non-commutative Khintchine type inequalities), a non-uniform recovery result with a linear estimate of $M$ in $K$ up to log-factors has been proven by Rauhut~\cite{rauhut9circulant}. For a detailed description of non-uniform recovery and its comparison to uniform recovery, one could refer to a paper by Rauhut~[Sections 3.1 and 4.2,~\cite{rauhut2010compressive}]. The behavior of $\rho\left(\vc{a}\right)$ also has important implications in the binary detection problem which we discuss in the next section.


\section{Compressive Binary Detection}
\label{sec:CBD}

\subsection{Problem Setup}

In this section, we address the problem of detecting a change in the dynamics of a linear system. We aim to perform the detection from the smallest number of observations, and for this reason, we call this problem Compressive Binary Detection (CBD).

\begin{figure}[tb]
\begin{center}
\includegraphics[width = 0.6\columnwidth]{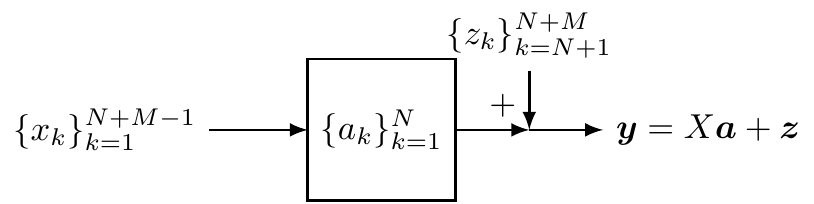}
\end{center}
\caption{\small\sl \ac{FIR} filter of order $N$ with impulse response $\left\{a_k\right\}_{k=1}^{N}$.}
\label{FIRmodel1}
\end{figure}

We consider an \ac{FIR} filter with a known impulse response $\vc{a} = \{a_{k}\}_{k=1}^{N}$. The response of this filter to a test signal $\vc{x} = \{x_{k}\}_{k=1}^{N+M-1}$ is described in~(\ref{firfilter}). We suppose the output of this filter is corrupted by random additive measurement noise $\vc{z}$. Fig.~\ref{FIRmodel1} shows the schematic of this measurement process.

From a collection of $M$ measurements $\vc{y}$ with $M < N$, our specific goal is to detect whether the dynamics of the system have changed to a different impulse response $\vc{b} = \left\{b_{k}\right\}_{k=1}^{N}$, which we also assume to be known. Since the the nominal impulse response $\vc{a}$ is known, the expected response $X\vc{a}$ can be subtracted off from $\vc{y}$, and thus without loss of generality, our detection problem can be stated as follows~\cite{csp}: Distinguish between two events which we define as $\mathcal{E}_0 \triangleq \left\{\vc{y} = \vc{z}\right\}$ and $\mathcal{E}_1 \triangleq \left\{\vc{y} = X\vc{c}+\vc{z}\right\}$, where $\vc{c} = \vc{b} - \vc{a}$ and $\vc{z}$ is a vector of \ac{i.i.d.} Gaussian noise with variance $\sigma^2$.

For any detection algorithm, one can define the false-alarm probability $P_{FA} \triangleq \Prob{\left(\mathcal{E}_1 \ \text{chosen when} \ \mathcal{E}_0\right)}$ and the detection probability $P_D \triangleq \Prob{\left(\mathcal{E}_1 \ \text{chosen when} \ \mathcal{E}_1\right)}$.
A \ac{ROC} is a plot of $P_D$ as a function of $P_{FA}$. A \ac{NP} detector maximizes $P_D$ for a given limit on the failure probability, $P_{FA} \leq \alpha$. The \ac{NP} test for our problem can be written as
$
\vc{y}^TX\vc{c} \overset{\mathcal{E}_1}{\underset{\mathcal{E}_0}{\gtrless}} \gamma,
$
where the threshold $\gamma$ is chosen to meet the constraint $P_{FA} \leq \alpha$. Consequently, we consider the detection statistic $d := \vc{y}^TX\vc{c}$.
By evaluating $d$ and comparing to the threshold $\gamma$, we are now able to decide between the two events $\mathcal{E}_0$ and $\mathcal{E}_1$. To fix the failure limit, we set $P_{FA} = \alpha$ which leads to
\begin{equation}
P_D\left(\alpha\right) = Q\left(Q^{-1}(\alpha)-\frac{\|X \vc{c} \|_2}{\sigma}\right),
\label{eq:ROC}
\end{equation}
where $Q(q) = \frac{1}{\sqrt{2 \pi}}\int_q^{\infty} e^{-\frac{u^2}{2}}\,du.$
As is evident from~(\ref{eq:ROC}), for a given $\alpha$, 
$P_D\left(\alpha\right)$ directly depends on $\|X\vc{c}\|_2$. On the other hand, because $X$ is a compressive random Toeplitz matrix, Theorem~\ref{maintheo1} suggests that $\|X\vc{c}\|_2$ is concentrated around its expected value with high probability and with a tail probability bound that decays exponentially in $M$ divided by $\rho\left(\vc{c}\right)$. Consequently, one could conclude that for fixed $M$, the behavior of $\rho\left(\vc{c}\right)$ affects the behavior of $P_D\left(\alpha\right)$ over $\alpha$. The following example illustrates this dependency.

\begin{example} (Detector Performance)
Assume with a failure probability of $P_{FA}=\alpha = 0.05$, a detection probability of $P_D\left(\alpha\right) = 0.95$ is desired. Assume $\sigma = 0.3$. From (\ref{eq:ROC}) and noting that $Q\left(-1.6449\right) = 0.95$ and $Q^{-1}\left(0.05\right) = 1.6449$, one concludes that in order to achieve the desired detection, $\|X\vc{c}\|_2$ should exceed $0.3\times 2\times 1.6449 = 0.9869$ (i.e., $\|X\vc{c}\|_2^2 \geq 0.9741$). On the other hand, for a Toeplitz $X$ with \ac{i.i.d.} entries drawn from $\mathcal{N}\left(0, \frac{1}{M}\right)$, $\E{\|X\vc{c}\|_2^2} = \|\vc{c}\|_2^2$. 
Assume without loss of generality, $\|\vc{c}\|_2 = 1$. Thus, from Theorem~\ref{maintheo1} and the bound in~(\ref{eq2}), we have for $\epsilon \in \left(0,1\right)$
\begin{equation}
\Prob{\|X\vc{c}\|_2^{2} - 1 \leq -\epsilon } \leq e^{-\frac{\epsilon^{2}M}{8\mu\left(\vc{c}\right)}} \leq e^{-\frac{\epsilon^{2}M}{8\rho\left(\vc{c}\right)}}.
\label{eq:detect_perform}
\end{equation}
Therefore, for a choice of $\epsilon = 1-0.9741 = 0.0259$ and from~(\ref{eq:detect_perform}), one could conclude that 
\[
\Prob{\|X\vc{c}\|_2^{2}  \leq 0.9741} \leq e^{-\frac{6.7\times 10^{-4}M}{8\rho\left(\vc{c}\right)}}.
\]
Consequently, for $\zeta \in \left(0,1\right)$, if $M \geq \frac{16\rho\left(\vc{c}\right)}{6.7\times 10^{-4}}\log{\zeta^{-1}}$, then with probability at least $1-\zeta^2$, $\|X\vc{c}\|_2^2$ exceeds $0.9741$, achieving the desired detection performance. \label{exm:detect_perform}
Apparently, $M$ depends on $\rho\left(\vc{c}\right)$ and qualitatively, one could conclude that for a fixed $M$, a signal $\vc{c}$ with small $\rho\left(\vc{c}\right)$ leads to better detection (i.e., maximized $P_D\left(\alpha\right)$ over $\alpha$). Similarly, a signal $\vc{c}$ with large $\rho\left(\vc{c}\right)$ is more difficult to reliably detect. 
\end{example}

In the next section, we examine signals of different $\rho\left(\vc{c}\right)$ values and show how their \acp{ROC} change. It is interesting to note that this dependence would not occur if the matrix $X$ were unstructured (which, of course, would not apply to the convolution-based measurement scenario considered here but is a useful comparison) as the \ac{CoM} behavior of unstructured Gaussian matrices is agnostic to the signal $\vc{c}$.

\begin{figure}[tb]
\centering
\subfigure[]{
   \includegraphics[width = .40\columnwidth]{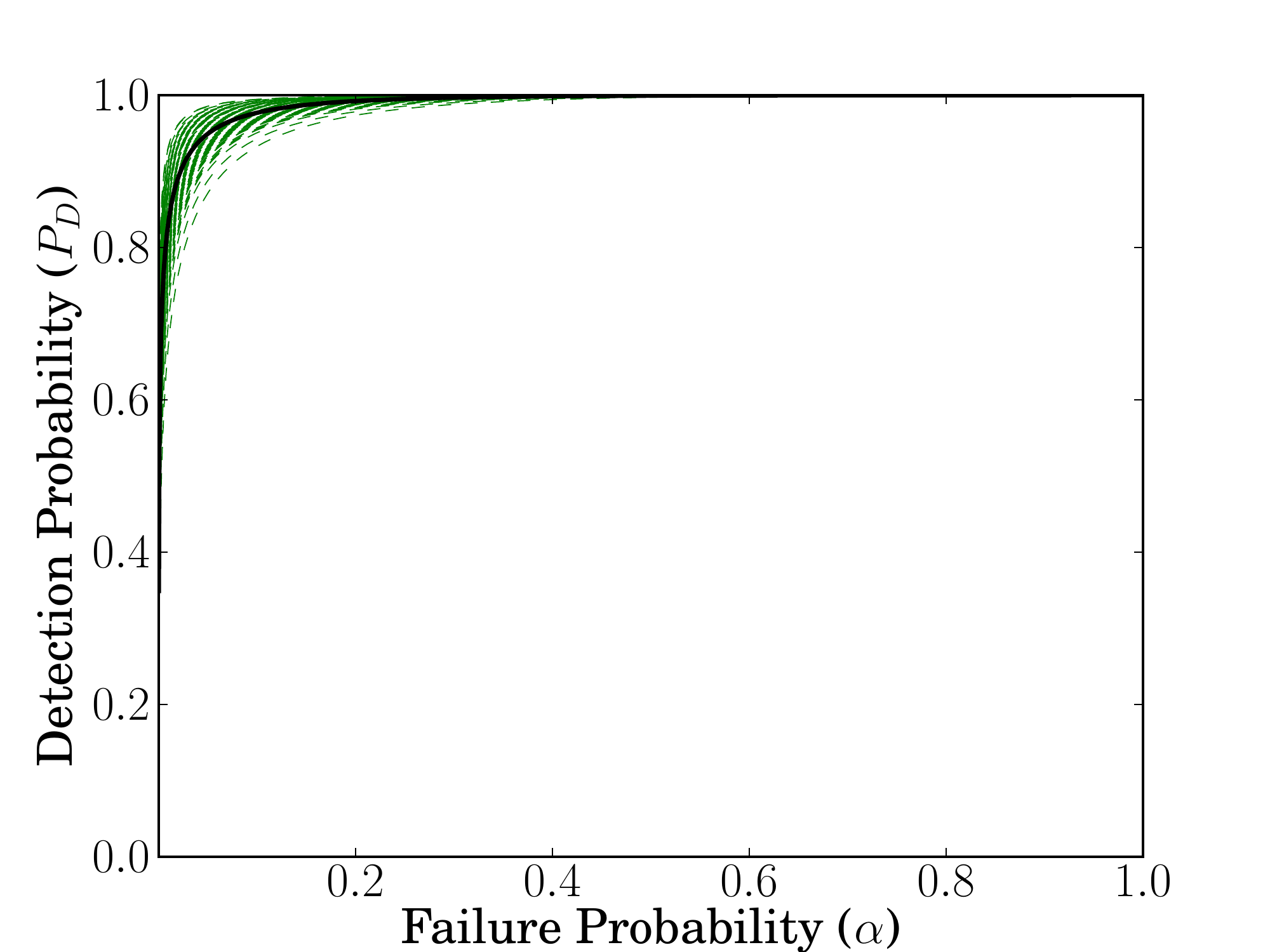}
 }
\subfigure[]{
   \includegraphics[width = .40\columnwidth]{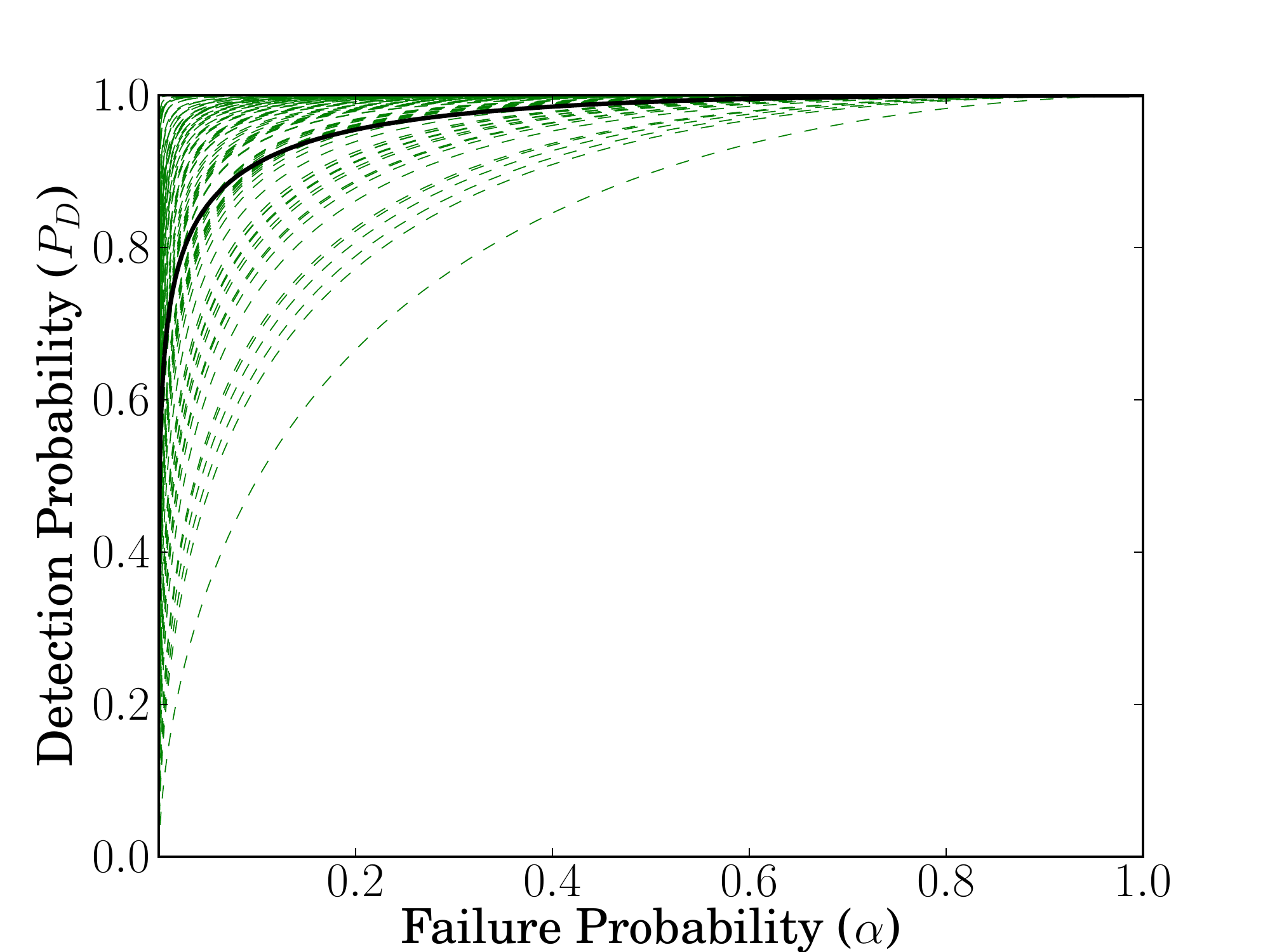}
 }
\caption{\small\sl ROCs for $1000$ random matrices $X$ for a fixed signal $\vc{c}$ with $\rho\left(\vc{c}\right) = 45.6$. (a) Unstructured $X$. (b) Toeplitz $X$. The solid black curve is the average of $1000$ curves.}
\label{fig:ROCcurve}
\end{figure}


\subsection{Empirical Results and \acp{ROC}}

In several simulations, we examine the impact of $\rho\left(\vc{c}\right)$ on the detector performance. To begin, we fix a signal $\vc{c} \in \real^{256}$ 
with $50$ non-zero entries all taking the same value; this signal has  $\|\vc{c}\|_2 = 1$ and $\rho\left(\vc{c}\right) = 45.6$ with our choice of $M = 128$. We generate $1000$ random unstructured and Toeplitz matrices $X$ with \ac{i.i.d.} entries drawn from $\mathcal{N}\left(0, \frac{1}{M}\right)$. For each matrix $X$, we compute a curve of $P_D$ over $P_{FA}$ using~(\ref{eq:ROC}); we set $\sigma = 0.3$. Figures~\ref{fig:ROCcurve}(a) and~\ref{fig:ROCcurve}(b) show the \acp{ROC} resulting from the unstructured and Toeplitz matrices, respectively.
As can be seen, the \acp{ROC} associated with Toeplitz matrices are more scattered than the \acp{ROC} associated with unstructured matrices. This is in fact due to the weaker concentration of $\norm{X\vc{c}}_2$ around its expected value for Toeplitz $X$ (recall (\ref{eq1}) and (\ref{eq2})) as compared to unstructured $X$ (recall (\ref{eq:conc_unstruct})).

\begin{figure}[tb]
\centering
\subfigure[]{
   \includegraphics[width = .40\columnwidth]{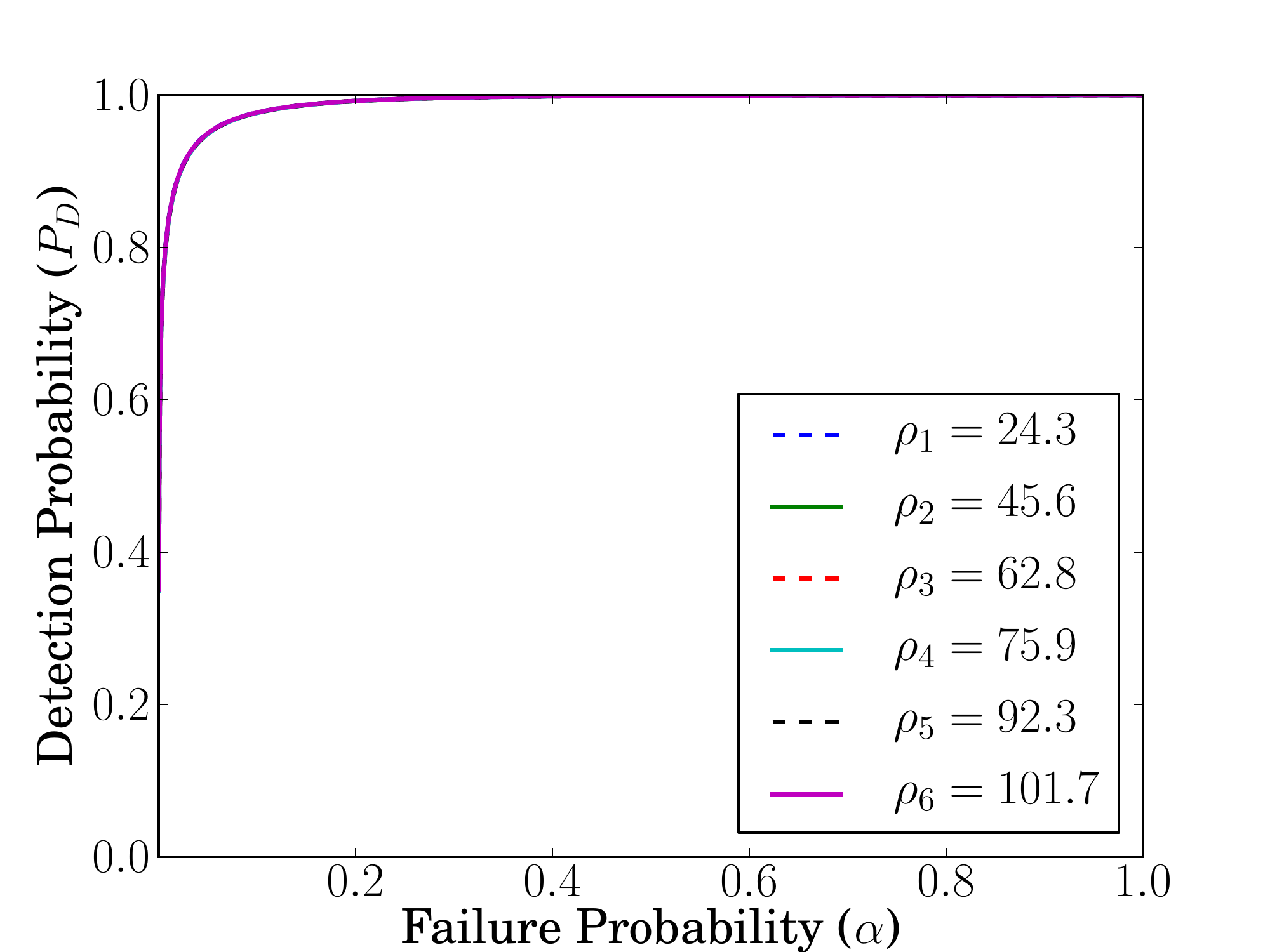}
   \label{fig:ROCcurveRandom}
 }
\subfigure[]{
   \includegraphics[width = .40\columnwidth]{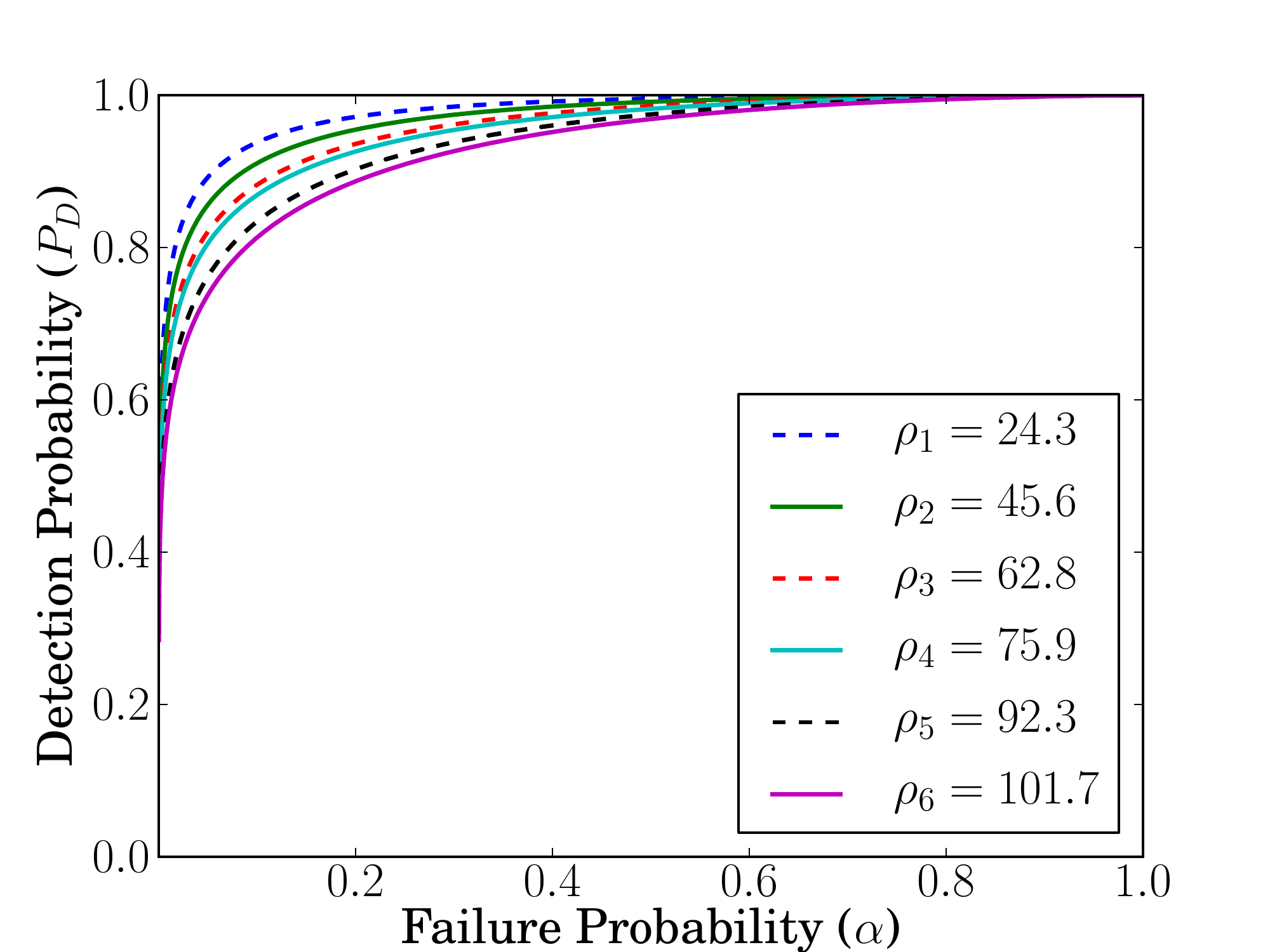}
   \label{fig:ROCcurveToep}
 }
\caption{\small\sl Average \acp{ROC} over $1000$ random matrices $X$ for $6$ different signals $\vc{c}$. (a) Unstructured $X$. All curves are overlapping. (b) Toeplitz $X$.
The curves descend in the same order they appear in legend box.}
\end{figure}

To compare the \acp{ROC} among signals having different $\rho\left(\vc{c}\right)$ values, we design a simulation with $6$ different signals. Each signal again has $\|\vc{c}\|_2 = 1$, and we take $\sigma = 0.3$ as above. Figures~\ref{fig:ROCcurveRandom} and \ref{fig:ROCcurveToep} plot the average \ac{ROC} for each signal over 1000 random unstructured and Toeplitz matrices, respectively. Two things are evident from these plots. First, the plots associated with Toeplitz matrices show a signal dependency while the ones associated with unstructured matrices are signal-agnostic. Second, with regards to the plots associated with Toeplitz $X$, we see a decrease in the curves (i.e., inferior detector performance) for signals with larger values of $\rho\left(\vc{c}\right)$.

In summary, our theory suggests and our simulations confirm that the value of $\rho\left(\vc{c}\right)$ has a direct influence on the detector performance. From a systems perspective, as an example, this means that detecting changes in systems having a sparse impulse response in the time domain (e.g., communication channels with multipath propagation) will be easier than doing so for systems having a sparse impulse response in the frequency domain (e.g., certain resonant systems). It is worth mentioning that while detection analysis of systems with sparse impulse response is interesting, our analysis can be applied to situations where neither the impulse responses $\vc{a}$ and $\vc{b}$ nor the change $\vc{c}$ are sparse.

\section*{Acknowledgment}

The authors gratefully acknowledge Chris Rozell, Han Lun Yap, Alejandro Weinstein, and Luis Tenorio for helpful conversations during the development of this work. The first author would like to thank Prof.\ Kameshwar Poolla and the Berkeley Center for Control and Identification at University of California at Berkeley for hosting him during Summer 2011; parts of this work were accomplished during that stay.

\appendix

\section{Proofs}

\subsection{Proof of Lemma~\ref{coro:prob_1}}
\label{app:coro:prob_1}

\prf{We start by proving a more general version of~(\ref{coro:prob_1_sum}). Let $z_1, z_2, \dots, z_n$ be random variables. Consider the event $\mathcal{E}_A \triangleq \left\{z_1 < c_1U \ \text{and} \ z_2 < c_2U \ \text{and} \ \cdots \ z_n < c_nU\right\}$ where $c_1, c_2, \dots, c_n$ are fixed numbers that sum to $1$. It is trivial to see that if $\mathcal{E}_A$ happens, then the event $\mathcal{E}_B \triangleq \left\{z_1 + z_2 + \cdots + z_n < U\right\}$ must also occur.
Consequently, $\Prob{(\mathcal{E}_{B})^c} \leq \Prob{(\mathcal{E}_{A})^c}$, where 
\[
(\mathcal{E}_{A})^c = \left\{z_1 \geq c_1U \ \text{or} \ z_2 \geq c_2U \ \text{or} \ \cdots \ z_n \geq c_nU\right\}.
\]
Using the union bound, we have $\Prob{(\mathcal{E}_{A})^c} \leq
\Prob{z_1 \geq c_1U} + \Prob{z_2 \geq c_2U} + \cdots + \Prob{z_n \geq c_nU}$ which completes the proof. The inequality~(\ref{coro:prob_1_sum}) is a special case of this result with $c_1 = c_2 = 0.5$.

We follow a similar approach for proving~(\ref{coro:prob_1_product}) where $z_1$ and $z_2$ are positive random variables. Consider the event $\mathcal{E}_A \triangleq \left\{z_1 < U_1 \ \text{and} \ z_2 > U_2\right\}$. If $\mathcal{E}_A$ occurs, then $\mathcal{E}_B \triangleq \left\{\frac{z_1}{z_2} < \frac{U_1}{U_2}\right\}$
must also occur. Consequently, $\Prob{(\mathcal{E}_{B})^c} \leq \Prob{(\mathcal{E}_{A})^c}$, where $(\mathcal{E}_{A})^c = \left\{z_1 \geq U_1 \ \text{or} \ z_2 \leq U_2\right\}$. Using the union bound, we have $\Prob{(\mathcal{E}_{A})^c} \leq \Prob{z_1 \geq U_1} + \Prob{z_2 \leq U_2}$.
}


\subsection{Proof of Proposition~\ref{propos:hoeff_complex_Gauss}}
\label{app:propos:hoeff_complex_Gauss}

Before proving Proposition~\ref{propos:hoeff_complex_Gauss}, we state the following lemma.
\begin{lemma}{(\emph{Hoeffding's inequality for real-valued Gaussian sums})}
Let $\vc{b} \in \mathbb{R}^N$ be fixed, and let $\vc{\epsilon} \in \mathbb{R}^N$ be a random vector whose $N$ entries are \ac{i.i.d.} random variables drawn from a Gaussian distribution with $\mathcal{N} (0, \sigma^2)$. Then, for any $u > 0$,
\begin{equation*}
\Prob{\left|\sum_{i=1}^{N}\epsilon_ib_i\right|\geq u} \leq e^{-\frac{u^2}{2\sigma^2\|\vc{b}\|_2^2}}.
\end{equation*}
\label{lemma:hoeff_real_Gauss}
\end{lemma}

\prf{
First note that the random variable $\sum_{i=1}^{N}\epsilon_ib_i$ is also Gaussian with distribution $\mathcal{N}\left(0, \sigma^2\|\vc{b}\|_2^2\right)$.
Applying a Gaussian tail bound to this distribution yields the inequality~\cite{tropp2007signal}. 
}

Using the result of Lemma~\ref{lemma:hoeff_real_Gauss}, we can complete the proof of Proposition~\ref{propos:hoeff_complex_Gauss}.

\prf{
Let $b_i = r_i + q_i j \ \forall i$ where $r_i$ is the real part of $b_i$ and $q_i$ is the imaginary part. Then we have
\begin{align*}
\Prob{\left|\sum_{i = 1}^{N}\epsilon_ib_i\right|\geq \|\vc{b}\|_2u}
&= \Prob{\left|\left(\sum_{i = 1}^{N}\epsilon_ir_i\right) + \left(\sum_{i = 1}^{N}\epsilon_iq_i\right) j\right|\geq \|\vc{b}\|_2u}\\
&= \Prob{\left|\left(\sum_{i = 1}^{N}\epsilon_ir_i\right) + \left(\sum_{i = 1}^{N}\epsilon_iq_i\right) j\right|^2\geq \|\vc{b}\|_2^2u^2}\\
&\leq \Prob{\left(\sum_{i = 1}^{N}\epsilon_ir_i\right)^2 \geq \frac{\|\vc{b}\|_2^2u^2}{2}} +
\Prob{\left(\sum_{i = 1}^{N}\epsilon_iq_i\right)^2 \geq \frac{\|\vc{b}\|_2^2u^2}{2}} \\
&= \Prob{\left|\sum_{i = 1}^{N}\epsilon_ir_i\right| \geq \frac{\|\vc{b}\|_2u}{\sqrt{2}}} +
\Prob{\left|\sum_{i = 1}^{N}\epsilon_iq_i\right| \geq \frac{\|\vc{b}\|_2u}{\sqrt{2}}}\\
&\leq e^{-\frac{\|\vc{b}\|_2^2u^2}{4\|\vc{r}\|_2^2\sigma^2}} + e^{-\frac{\|\vc{b}\|_2^2u^2}{4\|\vc{q}\|_2^2\sigma^2}}\leq 2e^{-\frac{u^2}{4\sigma^2}},
\end{align*}
where the first inequality uses Lemma~\ref{coro:prob_1} and the last inequality uses Lemma~\ref{lemma:hoeff_real_Gauss} and the facts that $\frac{\|\vc{b}\|_2}{\|\vc{r}\|_2} \geq 1$ and $\frac{\|\vc{b}\|_2}{\|\vc{q}\|_2} \geq 1$.
}

\bibliographystyle{IEEEtran}
\bibliography{referfile}

\end{document}